\documentclass[aoas,MSNbibl,nameyear,seceqn,dvips]{arximspdf}
\usepackage{graphicx}
\doi{10.1214/14-AOAS734} 
\volume{8}
\issue{3}
\pubyear{2014}
\firstpage{1314}
\lastpage{1340}

\makeatletter

\newcommand{\uI} {\mathbf{I}}
\newcommand{\uY} {\mathbf{Y}}
\newcommand{\udelta} {\bolds{\delta}}
\newcommand{\utheta} {\bolds{\theta}}
\newcommand{\utau} {\bolds{\tau}}
\makeatother

\begin{document}
\begin{frontmatter}

\title{A generalized mixed model framework for assessing
fingerprint individuality in presence of varying image
quality\thanksref{T1}}

\runtitle{A mixed model approach for individuality}
\thankstext{T1}{Supported in part by NSF Grant SES-0961649,
NSF Grant DMS-11-06450 and Universiti Teknologi PETRONAS URIF Grant no. 16/2013.}

\begin{aug}
\author[A]{\fnms{Sarat C.}~\snm{Dass}\corref{}\thanksref{TT1}\ead[label=e1]{saratcdass70@gmail.com}},
\author[B]{\fnms{Chae Young}~\snm{Lim}\thanksref{TT2}\ead[label=e2]{lim@stt.msu.edu}}
\and
\author[B]{\fnms{Tapabrata}~\snm{Maiti}\thanksref{TT2}\ead[label=e3]{maiti@stt.msu.edu}}
\runauthor{S.~C. Dass, C.~H. Lim and T. Maiti}
\affiliation{Universiti Teknologi PETRONAS\thanksmark{TT1}
and Michigan State
University\thanksmark{TT2}}
\address[A]{S. C. Dass \\
Department of Fundamental \& Applied Sciences\\
Universiti Teknologi PETRONAS\\
31750 Tronoh, Perak\\
Malaysia\\
\printead{e1}}
\address[B]{C. Y. Lim\\
T. Maiti\\
Department of Statistics \& Probability\\
Michigan State University\\
East Lansing, Michigan 48824\\
USA\\
\printead{e2}\\
\phantom{E-mail:\ }\printead*{e3}}
\end{aug}

\received{\smonth{7} \syear{2011}}
\revised{\smonth{2} \syear{2014}}

%
\begin{abstract}
Fingerprint individuality refers to the extent of uniqueness of
fingerprints and is the main criteria for deciding between a match
versus nonmatch in forensic testimony. Often, prints are subject to
varying levels of noise, for example, the image quality may be low when
a print is lifted from a crime scene. A poor image quality causes human
experts as well as automatic systems to make more errors in feature
detection by either missing true features or detecting spurious ones.
This error lowers the extent to which
one can claim individualization of fingerprints that are being matched.
The aim of this paper is to quantify the decrease in individualization
as image quality degrades based on fingerprint images in real
databases. This, in turn, can be used by forensic experts along with
their testimony in a court of law. An important practical concern is
that the databases used typically consist of a large number of
fingerprint images so computational algorithms such as the Gibbs
sampler can be extremely slow. We develop algorithms based on the
Laplace approximation of the likelihood and infer the unknown
parameters based on this approximate likelihood. Two publicly available
databases, namely, FVC2002 and FVC2006, are analyzed from which
estimates of individuality are obtained. From a statistical
perspective, the contribution can be treated as an innovative
application of Generalized Linear Mixed Models (GLMMs) to the field of
fingerprint-based authentication.
\end{abstract}

%
\end{frontmatter}

\section{Introduction}
\label{section:introduction}
\subsection{Fingerprint features, matching and individuality}
\begin{figure}

\includegraphics{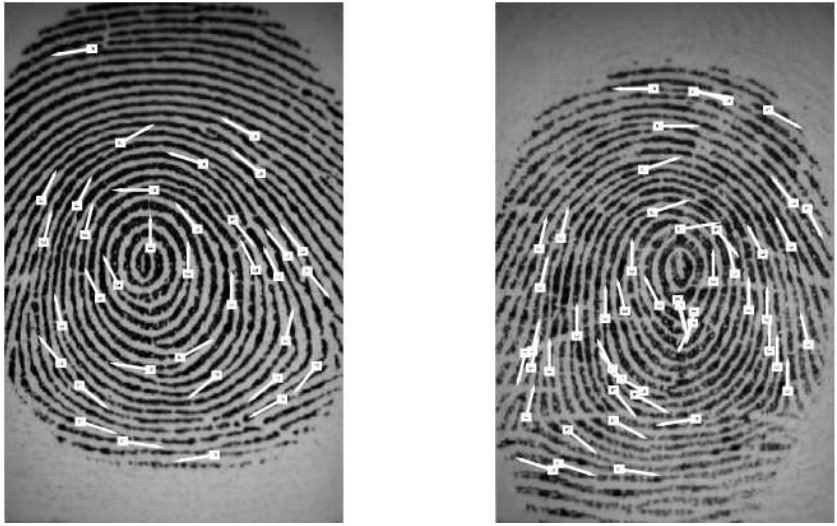}

\caption{A total of $35$ and $49$
minutiae were detected in left and right images, respectively, and
$w=7$ correspondences (i.e., matches) were found. The white squares
and lines, respectively, represent the minutia location and direction.
Note that for a minutia ending, the white line points away from the
ridge flow, whereas for a minutia bifurcation, the white line points
along the direction of the merged ridge. Images taken from the publicly
available database FVC2002 DB2.} \label{fig:impostor:match}
\end{figure}

Fingerprint individuality is the
study of the extent of uniqueness of fingerprints and is the central
premise for expert testimony in court. Assessment of individuality is
made based on comparing features from two fingerprint images such as in
Figure~\ref{fig:impostor:match}. A fingerprint image consists of
alternating dark and light smooth flow lines termed as ridges and
valleys. The fingerprint feature formed when a ridge occasionally
either terminates or bifurcates is called a minutia; the minutia type
``ending'' and ``bifurcation'' correspondingly relate to the type of
ridge anomaly that occurs. The location of a minutia is denoted by $x$
where $x \in D$, $D$ is the image domain. Figure~\ref{fig:impostor:match} identifies all minutiae as white squares in the
two fingerprint images; the images are from the publicly available
database FVC2002. Each white line emanating from the center of the
square represents the minutia direction, denoted by $u$ with $u \in
(0,2\pi]$, which is (i) the direction of the merged ridge flow for a
minutia bifurcation, and (ii) the direction pointing away from the
ridge flow for a minutia ending; see Figure~\ref{fig:impostor:match}.
Minutia information of a fingerprint consists of the collection of
locations and directions, $(x,u)$, of all minutiae in the image.

Minutiae in fingerprint images are extracted using pattern recognition
algorithms. In this paper, we used the algorithm described in Zhu, Dass and Jain (\citeyear{ZDJ07}) for minutia extraction. Minutia information is easy to
extract, and believed to be permanent and unique; that is, minutiae information
is believed to stay the same over time and different individuals have
distinct minutia patterns, thus making it a popular method for
identifying individuals in the forensics community. For a pair of
prints, a minutia $(x,u)$ in one print is said to match a minutia
$(y,v)$ in the another print if
%
\begin{equation}
\label{distances} |x-y|_{d} < r_0 \quad\mbox{and}\quad
|u-v|_{a}< u_0
\end{equation}
for prespecified small positive numbers $r_0$ and $u_0$, where
$|x-y|_{d}$ and $|u-v|_{a}$, respectively, denote the Euclidean
distance in $R^{2}$ and the angular distance
%
\begin{equation}
\label{angulardistance} |u-v|_{a} =
\operatorname{min}\bigl\{|u-v|,2\pi-|u-v|\bigr\}.
\end{equation}
We subsequently assume that a set of minutiae has already been
extracted (or detected) for every fingerprint image under study.
Fingerprint-based authentication proceeds by determining the highest
possible number of minutia matches between a pair of prints. This is
achieved by an optimal rigid transformation that brings the two sets of
minutiae as close to each other as possible and then counting the
number of minutia pairs $(x,u)$ and $(y,v)$
that satisfy (\ref{distances}). For example, the number of minutia
matches between the prints in the left and right panels in Figure~\ref{fig:impostor:match} is $w \equiv7$.

A reasonably high degree of match (high $w$) between the two sets of
minutiae leads forensic experts to testify irrefutably that the owner
of the two prints is one and the same person. Central to establishing
an identity based on fingerprint evidence is the assumption of
discernible uniqueness; fingerprint minutiae of different individuals
are observably different and, therefore, when two prints share many
common minutiae, the experts conclude that the two different prints are
from the same person. The primary concern here is what constitutes a
``reasonably high degree of match?'' For example, in Figure~\ref{fig:impostor:match}, is $w=7$ large enough to conclude that the two
prints come from the same person? When fingerprint evidence is
presented in a court, the testimony of experts is almost always
included and, on cross-examination,
the foundations and basis of this testimony are rarely
questioned. However, in the case of Daubert vs. Merrell Dow
Pharmaceuticals (\citeyear{Daubert}), 
the U.S. Supreme Court ruled
that in order for expert forensic testimony to be allowed in
courts, it had to be subject to the criteria of scientific
validation [see Pankanti, Prabhakar and Jain (\citeyear{PPJ02}) for
details]. Following the Daubert ruling, 
forensic evidence based on fingerprints was first challenged in
the case of U.S. vs. Byron C. Mitchell (\citeyear{USvsByronCMitchell}) 
and, subsequently, in 20 other cases involving fingerprint
evidence. The main concern with an expert's testimony as well as
admissibility of fingerprint evidence is the problem of
individualization. The
fundamental premise for asserting the extent to which the prints match
each other (i.e., the extent of uniqueness of fingerprints) has not
been scientifically validated and matching error rates are
unknown [Pankanti, Prabhakar and Jain (\citeyear{PPJ02}), Zhu, Dass and Jain (\citeyear{ZDJ07})]; see also the
National Academy of Sciences (NAS) report (\citeyear{NAS:2009}). %

The central question in a court of law is ``What is the uncertainty
associated with the experts' judgement when matches are decided by
fingerprint evidence?'' How likely can an erroneous decision be made
for the given latent print? The main issue with expert testimony is the
lack of quantification of this uncertainty in the decision. To address these
concerns, several research investigations have proposed measures that
characterize the extent of fingerprint individuality; see Pankanti, Prabhakar and Jain (\citeyear{PPJ02}), 
Zhu, Dass and Jain (\citeyear{ZDJ07}) and the references therein. The primary aim of these
measures is to capture the inherent variability and uncertainty (in the
expert's assessment of a ``match'') when
an individual is identified based on fingerprint evidence.

A measure of fingerprint individuality is given by the probability of a
random correspondence (PRC), which is the probability that two sets of
minutiae, one from the query containing $m_1$ minutiae and the other
from the template fingerprint containing $m_2$ minutiae, randomly
correspond to each other with at least $w$ matches. Since large
(resp., small) $w$ is a measure of the extent of similarity
(resp., dissimilarity) between a pair of fingerprints, the PRC
should be a decreasing function of~$w$. Mathematically, the PRC
corresponding to $w$ matches is given by
%
\begin{equation}
\label{PRC} \operatorname{PRC}(w | m_1, m_2) = P(\mathcal{S} \ge w |
m_1, m_2),
\end{equation}
where the random variable $\mathcal{S}$ denotes the number of minutia
matches that result when two arbitrary fingerprints from a target
population are paired with each other; the notation used in (\ref{PRC})
also emphasizes the dependence of the PRC on the total number of
minutiae detected in the two prints, that is, $m_1$ and $m_2$. It is
clear from the above formula that $\operatorname{PRC}(w | m_1, m_2)$ decreases as
$w$ increases for fixed $m_1$ and~$m_2$. The distribution of $\mathcal
{S}$ is governed by the extent to which minutia matches occur randomly,
an unknown quantity which has been modeled in a variety of ways in the
literature. The main aim of research in fingerprint individuality is to
obtain reliable inference on the PRC, which is also unknown as a result
of the unknown extent of random minutia matches.

\subsection{Minutia matching models}
\label{subsection:databasevalidation}

Several fingerprint minutia matching\break models and expressions for the PRC
have been developed from a completely theoretical perspective [see
Pankanti, Prabhakar and Jain (\citeyear{PPJ02}) for a detailed account of these models]. These
models are distributions elicited for minutia occurrences, that is, the
distributions that arise from viewing each minutia $(x,u)$ as a random
occurrence in the fingerprint, occurring independently of each other. A
major drawback of these works is that the models for minutia
occurrences, and consequently matching probabilities and PRCs, are not
validated based on actual fingerprint images. This effort was first
carried out in Pankanti, Prabhakar and Jain (\citeyear{PPJ02}) based on real fingerprint
databases, albeit for a simple minutia matching model. Subsequent
improvements on validating matching models based on actual fingerprint
databases have been carried out in Zhu, Dass and Jain (\citeyear{ZDJ07}) and several other
works. Zhu, Dass and Jain (\citeyear{ZDJ07}), for example, demonstrated that when the
number of minutiae in the prints are moderate to large, the
distribution of $\mathcal{S}$ in (\ref{PRC}) can be approximated by a
Poisson distribution with mean $\lambda$, the expected number of random
matches, given by
%
\begin{equation}
\label{lambdaf1f2} \lambda=\lambda(f_1,f_2)=
m_1 m_2 \biggl( 2\pi r_{0}^{2}
u_{0} \int_{D} \int_{(0,2\pi]}
f_{1}(x,u) f_{2}(x,u) \,dx \,du \biggr),
\end{equation}
based on the minutia distributions $f_1$ and $f_2$ for prints $1$ and
$2$, respectively; in~(\ref{lambdaf1f2}), $m_1$ and $m_2$ are,
respectively, the number of minutiae in prints $1$ and $2$, and $r_0$
and $u_0$ are as in (\ref{distances}). The above formula for the mean
number of random matches can be understood in the following way: The
total number of pairings available for random matching from $m_1$
minutiae in print $1$ and $m_2$ minutiae in print $2$ is $m_1 m_2$.
The expression $2\pi r_{0}^{2} u_{0} \int_{D} \int_{(0,2\pi]}
f_{1}(x,u) f_{2}(x,u) \,dx \,du$ represents the probability of a single
random match when $r_0$ and $u_0$ are small compared to the image size.
It follows from the expected value of a binomial distribution that
$\lambda$ in (\ref{lambdaf1f2}) is the mean number of successes (i.e.,
random matches). The contribution of Zhu, Dass and Jain (\citeyear{ZDJ07}) was to
characterize the distribution of $\mathcal{S}$ (the extent of random
matching in the target population) by a single number, namely, the
probability of a (single) random match between two minutiae, one from
each of the prints in question. Since the probability of a random match
%
\begin{equation}
\label{singlematch} p(f_1,f_2) = 2\pi r_{0}^{2}
u_{0} \int_{D} \int_{(0,2\pi]}
f_{1}(x,u) f_{2}(x,u) \,dx \,du
\end{equation}
depends on the minutia distributions $f_1$ and $f_2$, Zhu, Dass and Jain (\citeyear{ZDJ07})
inferred these distributions from actual fingerprint databases by
representing them as a finite mixture of normals.

\begin{figure}

\includegraphics{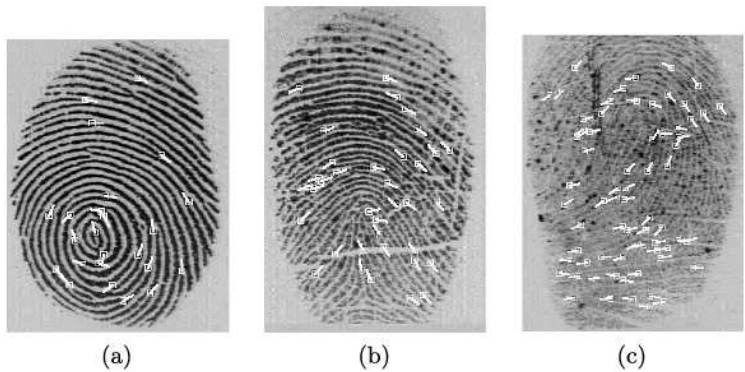}

\caption{Fingerprint images with \textup{(a)} good \textup{(b)}
moderate and \textup{(c)} poor quality. White squares and lines indicate
locations and directions of detected minutiae, and possibly not true
ones in panel \textup{(c)}; images taken from the publicly available FVC2002
databases.} \label{fig:imagequality}
\end{figure}

Factors other than the minutia distributions also govern the extent of
random matches. In the case of forensic testimony, for example, it is
reasonable to believe that expert matching is more prone to error if
the latent prints are of poor quality. Here, poor quality images mean
poor resolution (or clarity) of the ridge-valley structures (e.g., in the presence of smudges, sweaty fingers, cuts and bruises)
due to which the detection of true minutiae can be missed and spurious
minutiae can be detected; see Figure~\ref{fig:imagequality} for
examples of good, moderate and poor quality fingerprint images. In
other words, the extent of individualization should be smaller when the
underlying image qualities are poor. One critical issue, therefore, is
to be able to quantify the increase in PRC with respect to quality
degradation for each matching number $w$. This quantification is
crucial, for example, for latent prints lifted\vadjust{\goodbreak} from crime scenes which
are known to have inferior image quality. In the presence of poor quality,
automatic as well as manual extraction of fingerprint features are
prone to more errors through (i)~increased likelihood of detecting
spurious minutiae and (ii)~missing true ones. Previous work has not
quantified the effects of (i) and (ii) on fingerprint
individuality; see \citet{Dass10}.

\subsection{Objective and contributions}
The aim of this paper is to quantify PRC assessment (and hence
fingerprint individuality) in the presence of varying image quality.
The methodology involves two main steps. (i) First, a class of
generalized linear mixed models (GLMMs) is proposed that consists of
two levels. At the top level, the Poisson model is elicited as the
distribution for the number of random minutia matches, following the
derivation of Zhu, Dass and Jain (\citeyear{ZDJ07}). At the second level, image quality of
the two prints in question are incorporated as covariates. Also,
inter-finger variability is modeled using variance components in the
second level. (ii) Second, an inference procedure for the PRC is
developed to accommodate a large number of images from different
fingers as is typical in real fingerprint databases. Efficient
computational algorithms are essential for arriving at reliable reports
of fingerprint individuality in practice. Since the number of
fingerprint images in real databases is typically large, inference
based on Bayesian computational algorithms such as the Gibbs sampler is
extremely slow due to increased dimensions of the parameter space. To
alleviate this problem, the asymptotic Laplace approximation of the
GLMM likelihood is used instead. To summarize, \textit{contributions of
this paper are as follows}: For fingerprint-based authentication, a
procedure is developed for obtaining the PRC based on explicitly
modeling the occurrence of spurious minutia on image quality
degradation. Statistical contributions include (i) an innovative
application of GLMM to fingerprint-based authentication, and (ii) the
development of computationally fast algorithms achieved by
approximating the GLMM likelihood (with associated theoretical and
numerical validation). Inference results (point estimates and credible
intervals) on the PRC are also given (see the experimental results
section---Section~\ref{section:realdataanalysis}).

The rest of this paper is organized as follows: Section~\ref{section:glmm} discusses the
log-linear GLMM model that is used to study how PRCs change
as a function of the underlying image quality. Section~\ref{section:BayesianInference} presents the procedure for fitting
GLMMs in a Bayesian framework. Section~\ref{section:PRC} develops the
inference procedure (point estimates and credible intervals) on the
PRC. Section~\ref{section:realdataanalysis} obtains the estimates of
fingerprint individuality based on
the FVC2002 and FVC2006 databases and gives an analysis of the
numerical results. Section~\ref{simulation:results} validates the
Laplace approximation and various other approximations used in this
paper for efficient computation. Section~\ref{section:discussion} presents the summary, conclusions and other
relevant discussions.\vadjust{\goodbreak}

\section{Fingerprint databases and empirical findings}\label{sec2}

\subsection{Fingerprint databases} 
\begin{figure}

\includegraphics{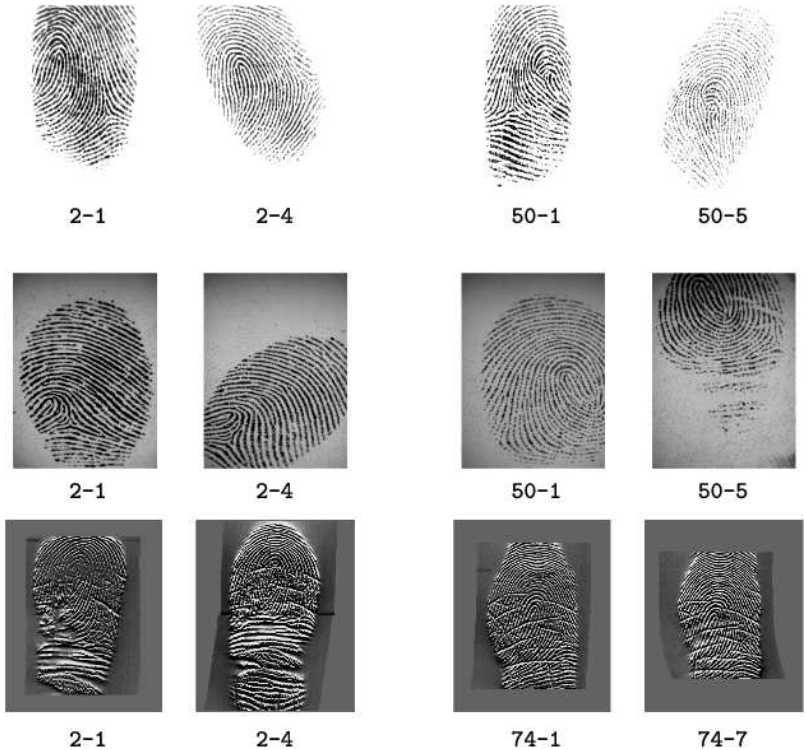}

\caption{Sample images from the FVC2002 DB1,
DB2 and FVC2006 DB3 databases; impression $f\mbox{--}l$ refers to the $l$th
impression of the $f$th finger. Top, middle and bottom rows are images
from the FVC2002 DB1, DB2 and FVC2006 DB3 databases, respectively; all
images are publicly available.} \label{fig:variousimages}
\end{figure}

Fingerprint databases that have been
used in fingerprint analysis include the FVC (e.g., FVC2000, 2002, 2004
and 2006) and NIST databases (e.g., NIST Special Database 4, 9, 27,
etc.) which are publicly available for download and analysis. 
These fingerprint databases typically consist of images acquired from
$F$ different fingers, obtained by placing the finger onto the sensing
plate of an image acquisition device. Each finger is sensed multiple
times, possibly a different number of times for the different fingers.
The databases have a common number of multiple acquisitions, say, $L$,
for each finger, resulting in total of $F L$ images in the database.
Due to the different sensors used as well as varying placement of the
finger, per individual, onto the sensing plate, the images acquired
exhibit significant variability (even for the same finger); see Figure~\ref{fig:variousimages} for several examples from two of the databases
used in this paper, namely, FVC2002 (DB1 and DB2) and FVC2006 (DB3).
For FVC2002 DB1 and DB2, $F=100$ and $L=8$, whereas for FVC2006 DB3,
$F=150$ with $L=15$. Figure~\ref{fig:variousimages} gives an idea of
the intra-finger variability (variability due to multiple impressions)
that has to be taken into account in addition to image quality
variability. Intra-finger variability gives different matching numbers
when different impressions are used for the matching between two fingers.

\subsection{Empirical findings for varying image quality}
\label{empiricalfindings}
In practice, the quality of an image can be either ordinal (i.e.,
taking values in an ordered label set) or quantitative (i.e., taking
values in a continuum, usually in a bounded interval $[a,b]$). We have
considered one specific choice of each type of quality measure: (i) The
categorical quality extractor ``NFIQ,'' which is an implementation of
the ``NIST Image Quality'' algorithm based on neural networks described
in Tabassi, Wilson and Watson (\citeyear{TWW04}), and (ii) the minutia quality extractor
obtained from the feature extraction algorithm \texttt{mindtct} from NIST
(see the Home Office Automatic Fingerprint Recognition System [HOAFRS (\citeyear{LIC93})]). These algorithms are all publicly available. For the NFIQ
quality extractor, the output $Q_0$ is a quality label numbered
$1,2,\ldots,5$, with $1$ and $5$ corresponding to the best and worst
quality images, respectively. For the continuous quality extractor by
\texttt{mindtct}, a real number $Q_{\mathrm{con}}$ between $0$ and $1$ is obtained
with higher values indicating better quality images. To maintain
consistency between the two quality measures, we relabel the categorial
measure as $Q_{\mathrm{cat}} = Q_{\mathrm{max}} + 1 - Q_{0}$, where $Q_{\mathrm{max}}$ is the
maximum label for a given database, so that higher labels indicate
better quality images.

For a pair of prints, $1$ and $2$ say, with $m_1$ and $m_2$ minutiae,
respectively, let $Y$ denote the observed number of minutia matches
between them. We consider only impostor matches (matches between
impressions of different fingers) which are equivalent to the notion of
random matching when assessing fingerprint individuality. The total
number of impostor pairs in a database with $F$ fingers and $L$
impressions per finger is $F(F-1)L^{2}/2$ (since the order of the
prints is immaterial). Suppose $(Q_1,Q_2)$ are the labels for the
categorical measure $Q_{{\mathrm{cat}}}$. Table~\ref{tableYFVC2002} gives the
average number (and standard deviations) of matches $\bar{Y}$ for each
quality bin pair based on FVC2002 DB1 and DB2. Note that the average
number of matches is an increasing function of the quality labels, that
is, the average number of random matches increases as the image quality
becomes better. This seems counterintuitive initially, but we note that
the average increases because the total number of minutiae extracted
from the two prints also increases as quality gets better. With more
minutiae available for random pairings, the number of matches based on
such pairings should also increase on the average. Table~\ref{tablemnFVC2002} gives the mean number (and standard deviations) of
minutiae for each quality bin for the two databases considered to
illustrate this trend. The minutia extraction algorithm \texttt{mindtct}
detects (or extracts) a minutia only if its computed reliability
measure (computed within the program) is above a certain threshold. For
better quality images, more minutiae have a reliability index above the
threshold, which explains the higher number extracted in better quality images.

\begin{table}
\caption{Mean and standard deviations (in
parenthesis) of $Y$ (the observed number of random matches) for
$Q_{\mathrm{cat}}$ pairs for FVC2002 DB1 and DB2 in panels \textup{(a)} and \textup{(b)}}
\label{tableYFVC2002}
\centering
\begin{tabular}{@{}cc@{}}
\begin{tabular}{@{}lccc@{}}\\
\hline
\multicolumn{1}{@{}l}{$\bolds{Q_1/Q_2}$} &
\multicolumn{1}{c}{$\bolds{1}$} &
\multicolumn{1}{c}{$\bolds{2}$} &
\multicolumn{1}{c@{}}{$\bolds{3}$} \\
\hline
$1$ & 3.36 & 3.43 & 4.15 \\
& (1.11) & (1.14) & (1.36) \\
$2$ & 3.73 & 4.02 & 4.94 \\
& (1.13) & (1.19) & (1.41) \\
$3$ & 4.57 & 4.96 & 6.38 \\
& (1.40) & (1.43) & (1.71) \\
\hline
\end{tabular}
&
\begin{tabular}{@{}lcccc@{}}
\hline
\multicolumn{1}{@{}l}{$\bolds{Q_1/Q_2}$} &
\multicolumn{1}{c}{$\bolds{1}$} &
\multicolumn{1}{c}{$\bolds{2}$} &
\multicolumn{1}{c}{$\bolds{3}$} &
\multicolumn{1}{c@{}}{$\bolds{4}$} \\
\hline
$1$ & 2.20 & 3.02 & 3.36 & 3.76\\
& (0.42) & (0.76) & (0.90) & (1.00) \\
$2$ & 2.96 & 3.91 & 4.21 & 5.12 \\
& (0.75) & (1.30) & (1.39) & (1.71) \\
$3$ & 3.19 & 4.26 & 4.71 & 5.77 \\
& (0.82) & (1.26) & (1.32) & (1.57)\\
$4$ & 3.74 & 5.26 & 5.96 & 7.47 \\
& (0.99) & (1.51) & (1.58) & (1.91) \\
\hline\\
\end{tabular}\\
\textup{(a)}& \textup{(b)}
\end{tabular}
\end{table}

\begin{table}
\caption{Mean and standard deviations (in
parenthesis) of $m_1 m_2$ (the total number of possible random
pairings) for $Q_{\mathrm{cat}}$ pairs for FVC2002 DB1 and DB2 in
panels \textup{(a)} and \textup{(b)}}
\label{tablemnFVC2002}
\centering
\begin{tabular}{@{}cc@{}}
\begin{tabular}{@{}lccc@{}}\\\\
\hline
\multicolumn{1}{@{}l}{$\bolds{Q_1/Q_2}$} &
\multicolumn{1}{c}{$\bolds{1}$} &
\multicolumn{1}{c}{$\bolds{2}$} &
\multicolumn{1}{c@{}}{$\bolds{3}$} \\
\hline
$1$ & 357 & 403 & 641 \\
& (239) & (219) & (336) \\
$2$ & 490 & 594 & 947 \\
& (260) & (221) & (327) \\
$3$ & 768 & 921 & 1459 \\
& (383) & (301) & (440) \\
\hline
\end{tabular}
&
\begin{tabular}{@{}lcccc@{}}
\hline
\multicolumn{1}{@{}l}{$\bolds{Q_1/Q_2}$} &
\multicolumn{1}{c}{$\bolds{1}$} &
\multicolumn{1}{c}{$\bolds{2}$} &
\multicolumn{1}{c}{$\bolds{3}$} &
\multicolumn{1}{c@{}}{$\bolds{4}$} \\
\hline
$1$ & 177 & 366 & 482 & 705\\
& (53) & (144) & (148) & (203) \\
$2$ & 338 & 684 & 835 & 1295 \\
& (178) & (395) & (445) & (634) \\
$3$ & 414 & 845 & 1036 & 1596 \\
& (138) & (368) & (350) & (485)\\
$4$ & 650 & 1342 & 1682 & 2546 \\
& (192) & (527) & (491) & (655) \\
\hline
\end{tabular}\\
\textup{(a)}& \textup{(b)}
\end{tabular}
\end{table}

\begin{table}[b]
\tabcolsep=3pt
\caption{Mean and standard deviations (in
parenthesis) of $Y/(m_1 m_2)$ (the probability of a random pairing) for
$Q_{\mathrm{cat}}$ pairs for FVC2002 DB1 and DB2 in panels \textup{(a)} and \textup{(b)}}
\label{tablepFVC2002}
\centering
\begin{tabular}{@{}cc@{}}
\begin{tabular}{@{}lccc@{}}\\\\
\hline
\multicolumn{1}{@{}l}{$\bolds{Q_1/Q_2}$} &
\multicolumn{1}{c}{$\bolds{1}$} &
\multicolumn{1}{c}{$\bolds{2}$} &
\multicolumn{1}{c@{}}{$\bolds{3}$} \\
\hline
$1$ & 0.0127 & 0.0105 & 0.0077 \\
& (0.0083) & (0.0057) & (0.0036) \\
$2$ & 0.0092 & 0.0075 & 0.0056 \\
& (0.0046) & (0.0030) & (0.0019) \\
$3$ & 0.0069 & 0.0058 & 0.0046 \\
& (0.0029) & (0.0018) & (0.0013) \\
\hline
\end{tabular}
&
\begin{tabular}{@{}lcccc@{}}
\hline
\multicolumn{1}{@{}l}{$\bolds{Q_1/Q_2}$} &
\multicolumn{1}{c}{$\bolds{1}$} &
\multicolumn{1}{c}{$\bolds{2}$} &
\multicolumn{1}{c}{$\bolds{3}$} &
\multicolumn{1}{c@{}}{$\bolds{4}$} \\
\hline
$1$ & 0.0131 & 0.0090 & 0.0074 & 0.0056\\
& (0.0032) & (0.0029) & (0.0024) & (0.0021) \\
$2$ & 0.0104 & 0.0068 & 0.0058 & 0.0044 \\
& (0.0042) & (0.0028) & (0.0023) & (0.0017) \\
$3$ & 0.0083 & 0.0056 & 0.0049 & 0.0038 \\
& (0.0028) & (0.0021) & (0.0016) & (0.0012)\\
$4$ & 0.0060 & 0.0043 & 0.0037 & 0.0030 \\
& (0.0017) & (0.0014) & (0.0010) & (0.0009) \\
\hline
\end{tabular}\\
\textup{(a)}& \textup{(b)}
\end{tabular}
\end{table}

Consequently, a better quantity to model for varying image quality is
$Y/(m_1 m_2)$, which is an estimate of the probability of a random
match based on the Poisson model of Zhu, Dass and Jain (\citeyear{ZDJ07}) [see (\ref
{lambdaf1f2}) and (\ref{singlematch})]. Table~\ref{tablepFVC2002} gives
the average value of $Y/(m_1 m_2)$ (and standard deviations) for the
different quality bins. Note that now the averages are \textit{decreasing}
as a function of $Q_{\mathrm{cat}}$. Intuitively, this is expected since it is
less likely to obtain a random match when the image quality is high for
a pair of impostor fingerprints. More importantly, we can attribute the
larger values of $Y/(m_1 m_2)$ for poor quality images to the
extraction of spurious minutiae. The probability of a random match
based on true minutiae is intrinsic to the two prints in question and
depends only on the distributions $f_1$ and $f_2$. Thus, this
probability should not depend on quality if all true minutiae are
correctly identified. The probability of a random match, $p$, in the
presence of noisy (both spurious and true) minutiae is the sum of four
component probabilities:
%
\begin{equation}
\label{componentprobabilities} p = p^{(0,0)} + p^{(0,1)} + p^{(1,0)} +
p^{(1,1)},
\end{equation}
where $p^{(u,v)}$ is the probability of a random match with minutia of
type $u$ from print~$1$ and type $v$ from print $2$; $u=0,1$ for true
or spurious minutia from print~$1$, and $v=0,1$ for true or spurious
minutia from print $2$. From earlier discussion, it follows that
$p^{(0,0)}$ does not depend on $(Q_1,Q_2)$, whereas $p^{(0,1)},
p^{(1,0)}$ and $p^{(1,1)}$ all increase as either or both quality
labels $(Q_1,Q_2)$ decrease. In the GLMM framework, the dependence of
$p^{(0,1)}$, $p^{(1,0)}$ and $p^{(1,1)}$ on quality is modeled
explicitly. The above discussion is presented for the categorical
quality measure $Q_{\mathrm{cat}}$, but similar empirical findings are also
obtained for $Q_{\mathrm{con}}$ by binning the values in the range $[0,1]$.

\section{GLMM framework for fingerprint individuality}
\label{section:glmm}

Let a fingerprint\break database consists of $F$ fingers and $L$ impressions
per finger. Each fingerprint image in the database corresponds to a
fingerprint impression denoted by the index pair $(f,l)$ with $1\le f
\le F$ and $1 \le l \le L$; here $f$ and $l$, respectively, are the
indices of the finger and the impression. An impostor pair of
fingerprint images $(i,j)$ with $i\equiv(f,l)$ and $j \equiv(f',l')$
arise from different fingers, that is, $f\ne f'$. The collection of all
impostor pairs arising from the database is denoted by $\mathcal{I}$
with $N \equiv F(F-1) L^2/2$. For each impostor pair $(i,j) \equiv
((f,l),(f',l'))$, the matching algorithm presented in Section~\ref{section:introduction} [see (\ref{distances}) and (\ref
{angulardistance})] computes the observed number of minutia matches,
$Y_{ij}$, between impressions $i$ and $j$. The image quality of $i$ and
$j$ is obtained by a quality extractor $Q$ that outputs the ordered
pair $(Q_i,Q_j)$ which for the moment is taken to be continuous (the
categorical quality case is discussed later).

Based on the discussion in the previous section, the total number of
matches $Y_{ij}$ is expressed as a sum of four components:
%
\begin{equation}
\label{sumY} Y_{ij} = \sum_{u=0}^{1}
\sum_{v=0}^{1} Y_{ij}^{(u,v)},
\end{equation}
where $Y_{ij}^{(u,v)}$ is the number of matches obtained based on
minutia of type $u$ for impression $i$ and minutia of type $v$ for
impression $j$ [see (\ref{componentprobabilities}) and the ensuing
discussion in Section~\ref{empiricalfindings}]. The rest of the GLMM
model specification is
%
\begin{equation}
\label{simple:model:2} Y_{ij}^{(u,v)}
\sim \operatorname{Poisson} \bigl(\lambda
_{ij}^{(u,v)} \bigr),
\end{equation}
independently for each combination of $(u,v)$,
%
\begin{equation}
\label{mimj} \lambda_{ij}^{(u,v)} = m_i
m_j \operatorname{exp} \bigl\{ b_{f} + b_{f'} +
\eta_{ij}^{(u,v)} \bigr\}
\end{equation}
with $\eta_{ij}^{(u,v)}$ for various combinations $(u,v)$ given as
%
\begin{eqnarray}
\label{sim1} \eta_{ij}^{(0,0)} &=& 2\beta_0,
\\
\label{sim2} \eta_{ij}^{(0,1)} &=& \beta_0 +
\theta_0 + \theta_1 {Q_j},
\\
\label{sim3} \eta_{ij}^{(1,0)} &=& \beta_0 +
\theta_0 + \theta_1 Q_{i} \quad\mbox{and}
\\
\label{sim4} \eta_{ij}^{(1,1)} &=& 2\theta_0 +
\theta_1 (Q_i + Q_j).
\end{eqnarray}
In (\ref{mimj}), $m_i$ and $m_j$ are, respectively, the number of
extracted minutiae from $i$ and~$j$. The GLMM model of (\ref
{sumY})--(\ref{sim4}) consists of unknown fixed effects parameters
$(\theta_0,\theta_1,\beta_0)$ and random effect parameters $b_{f}$ with
$b_{f} \sim N(0,\sigma^{2})$ independently for $f=1,2,\ldots,F$ for
unknown variance $\sigma^{2}>0$.

Comparing (\ref{lambdaf1f2}) and (\ref{mimj}), we note that the
probability of a random match $p(f_1,f_2)$ [see (\ref{singlematch})] is
modeled as $\operatorname{exp}\{ b_{f} + b_{f'} + \eta_{ij}^{(u,v)}\}$ with
random effects $b_{f}$ and $b_{f'}$ in the GLMM framework. The reason
for this is as follows: The assessment of fingerprint individuality is
typically carried out for a target population with different
individuals. Hence, $f_1$ and $f_2$ are random realizations of prints
from the target population. While each $f_1$ and $f_2$ is modeled as a
mixture of normals, Zhu, Dass and Jain (\citeyear{ZDJ07}) subsequently proceed with a
clustering of these estimated $f_1$ and $f_2$ for a given database. The
assumption made is that the target population (and, hence, the database
which is considered a representative of the target population) consists
of unknown $K$ different clusters of hyperdistributions (a distribution
on the mixtures) from which $f_1$ and $f_2$ are realized. Subsequent
development of the clustering of mixtures of normals is reported in
\citet{D11} where the uncertainty of estimating the
hyperdistribution is accounted for in the assessment of PRC. In the
present context, the random effects $b_f$ (and $b_{f'}$) account for
variability due to different fingers (each finger $f$ has a
distribution on its minutiae) which are assumed to be realizations from
the target population. The fixed effect parameters $(\theta_0,\theta
_1,\beta_0,\sigma^{2})$ are target population specific: Their values
can change when we move from one database to another.

It follows that the parameters $(\theta_0,\theta_1,\beta_0)$ should be
all negative. More elaborate restrictions can be placed on the random
and fixed effects parameters jointly by requiring that $b_{f} + b_{f'}
+ \eta_{ij}^{(u,v)}\le0$ for all $(u,v)$ and $(i,j)$, but we choose
not to pursue estimation and subsequent Bayesian inference in this
complicated feasibility region. The posterior estimates of $(\theta
_0,\theta_1,\beta_0,\sigma^{2})$ obtained in the experimental results
section demonstrate that the restrictions are satisfied naturally:
Estimates of $(\theta_0,\theta_1,\beta_0)$ are found to be negative for
all the databases we worked with. Further, we found the estimate of
$\sigma$ to be so small relative to $(\theta_0,\theta_1,\beta_0)$ that
the restrictions $b_{f} + b_{f'} + \eta_{ij}^{(u,v)} \le0$ satisfied
automatically for all realized values of $b_f$ and $b_{f'}$ when
computing the PRC.

For a categorical quality measure labeled in increasing order of image
quality, (\ref{sim1})--(\ref{sim4}) in the GLMM framework are replaced
by the following four equations for $\eta_{ij}^{(u,v)}$:
%
\begin{eqnarray}
\label{sim11} \eta_{ij}^{(0,0)} &=& 2\beta_0,
\\
\label{sim21} \eta_{ij}^{(0,1)} &=& \beta_0 +
\theta_0 + \theta_1 +\cdots+\theta_{(Q_j - 1)},
\\
\label{sim31} \eta_{ij}^{(1,0)} &=& \beta_0 +
\theta_0 + \theta_1 + \cdots+\theta_{(Q_{i}-1)}\quad
\mbox{and}
\\
\label{sim41} \eta_{ij}^{(1,1)} &=& 2\theta_0 +
\theta_1 + \cdots+\theta_{(Q_i-1)} + \theta_1 +
\cdots+ \theta_{(Q_j-1)},
\end{eqnarray}
where the fixed effects parameters are now $(\theta_0,\theta_1,\ldots,\theta_{(Q_{\mathrm{max}}-1)})$ and $\beta_0$ which are all negative.

Equations (\ref{sim1})--(\ref{sim4}) for a continuous quality measure
and (\ref{sim11})--(\ref{sim41}) for a categorical quality measure imply
that the matching between type $u$ minutia from print $1$ and type $v$
minutia from print $2$ are independent of each other. To see this, note
that the probability of a random match in (\ref
{componentprobabilities}), $p=p^{(0,0)}+p^{(0,1)} + p^{(1,0)} +
p^{(1,1)}$, is the sum of four component probabilities depending on the
type of minutiae being matched. It follows that each normalized term
${p^{(u,v)}}/{p}$ gives the multinomial probability that type $u$
minutia is paired with type $v$ minutia, conditional on the fact that a
random pairing has occurred. For the GLMM framework, we get
${p^{(u,v)}}/{p} = {e^{\eta_{ij}^{(u,v)}}}/{\sum_{u=0}^{1}\sum_{v=0}^{1} e^{\eta_{ij}(u,v)}}$. Consequently, the odds ratio $\frac
{p^{(1,1)}p^{(0,0)}}{p^{(0,1)}p^{(1,0)}} = 1$ for both the continuous
and categorical quality measures. In other words, spurious minutia
locations are dispersed ``evenly'' in between the true minutiae. No
region in the print is more prone to spurious detection in relation to
the distribution of true minutiae in the fingerprint impression. We
further note that $\frac{p^{(u,1)}}{p^{(u,0)}} = e^{\theta_0 + \theta_1
Q_{j} - \beta_0}$ for the continuous and
$\frac{p^{(u,1)}}{p^{(u,0)}} = e^{\theta_0 + \theta_1 + \cdots+
\theta
_{(Q_{j}-1)} - \beta_0}$ for the categorical quality measures.
These expressions suggest that $\theta_1$ for the continuous and
$\theta
_1, \ldots, \theta_{(Q_{\mathrm{max}}-1)}$ for categorical quality measures
should all be negative: As $Q_{j}$ increases (better quality image),
the odds of pairing with spurious minutiae ($v=1$ versus $v=0$) should
decrease for each minutia type $u=0,1$.

For both the categorical and continuous quality measures, equation
(\ref
{mimj}) can be rewritten in the general log-linear form with respect to
the fixed and random effects parameters. For each fixed $(u,v)$, this
is given by
%
\begin{equation}
\label{random:effects:model} \operatorname{log} \lambda_{ij}^{(u,v)} =
K_{ij} + \delta_{ij}(u,v),
\end{equation}
where $K_{ij} = \operatorname{log}(m_i m_j)$ and $\delta_{ij}{(u,v)} =
x_{ij}'(u,v)\utheta+ z_{ij}'{\mathbf{b}}$
for appropriate\break choices of the $1 \times p$ row vector $x_{ij}'(u,v)$
and $1\times F$ row vector $z_{ij}'$ [which is independent of $(u,v)$];
$\utheta$ and $\mathbf{b} = (b_1,b_2,\ldots,b_{F})'$, respectively, are
$p \times1$ and $F \times1$ column vectors representing the
collection of fixed and random effects parameters. For the continuous
quality measure, the parameter vector ${\utheta}= (\theta_0,\theta
_1,\beta_0)'$ with $p=3$, whereas ${\utheta}= (\theta_0,\theta
_1,\ldots,\theta_{(Q_{\mathrm{max}}-1)},\beta_0)'$ with $p=Q_{\mathrm{max}}+1$ for the categorical
quality measure. We also denote $\utau= (\utheta',\sigma^{2})'$ to be
the $(p+1)\times1$ vector consisting of all unknown fixed effects parameters.

In matrix notation, the GLMM for each fixed $(u,v)$ is given by
%
\begin{equation}
\label{matrix:notation:0}\qquad Y_{ij}^{(u,v)} \sim \operatorname{Poisson}
\bigl( e^{K_{ij}+\delta
_{ij}{(u,v)}} \bigr)\qquad\mbox{independently for each $(i,j) \in
\mathcal{I}$}
\end{equation}
and
\begin{equation}
\label{matrix:notation} {\udelta} {(u,v)} = \mathbf{X}(u,v)\utheta+ \mathbf{Z}
\mathbf{b},
\end{equation}
where ${\udelta}{(u,v)}$ is the $N \times1$ vector consisting of
$\delta_{ij}{(u,v)}$s, $\mathbf{X}(u,v)$ is the $N \times p$ matrix
with rows consisting of $x_{ij}'(u,v)$, and $\mathbf{Z}$ is the $N
\times F$ matrix with rows comprising of $z_{ij}'$.

%

\section{Inference methodology}
\label{section:BayesianInference}

The subsequent subsections develop inference methodology for $\utau
\equiv(\utheta',\sigma^{2})'$ in a Bayesian framework.

\subsection{Exact GLMM likelihood}

The following notation is developed for the ensuing discussion. Let
$\mathbf{Y}_{uv} = \{ Y_{ij}^{(u,v)}\dvtx (i,j) \in\mathcal{I} \}$
denote the missing data component comprised of minutia matches of type
$(u,v)$ for all impostor pairs. We also denote $\mathbf{Y}_{\mathrm{mis}}$ and
$\mathbf{Y}_{\mathrm{obs}}$ to be the collection of all missing and observed
matching numbers, that is,
$\mathbf{Y}_{\mathrm{mis}} = ( \mathbf{Y}_{00}, \mathbf{Y}_{01}, \mathbf
{Y}_{10},\mathbf{Y}_{11} )$ and $\mathbf{Y}_{\mathrm{obs}} = \{ Y_{ij},
(i,j)\in\mathcal{I} \}$, respectively. The set of feasible values for
$\mathbf{Y}_{\mathrm{mis}}$ is given by the set
%
\begin{equation}
\label{missinglabel} \mathcal{M} = \Biggl\{ \mathbf{Y}_{\mathrm{mis}}\dvtx \sum
_{u=0}^{1}\sum_{v=0}^{1}
Y_{ij}^{(u,v)} = Y_{ij} \mbox{ for all } (i,j) \in
\mathcal {I} \Biggr\}.
\end{equation}
All subsequent analysis is conditional on the random effects $\mathbf
{b}$. The missing data component $\mathbf{Y}_{uv}$ has a likelihood
$\ell_{uv}(\mathbf{Y}_{uv} | \utheta,\mathbf{b}) =
e^{-h_{uv}(\mathbf
{Y}_{uv}, \utheta, \mathbf{b})}$,
where
%
\begin{eqnarray}
\label{huv} h_{uv}(\mathbf{Y}_{uv}, \utheta,
\mathbf{b})&=& - \sum_{(i,j)\in\mathcal{I}} \bigl\{ \bigl(
x_{ij}'(u,v)\utheta+ z_{ij}'
\mathbf{b}+K_{ij} \bigr) Y_{ij}^{(u,v)}
\nonumber
\\[-8pt]
\\[-8pt]
\nonumber
&&\hspace*{41pt}{} - \operatorname{exp} \bigl(x_{ij}'(u,v)\utheta+
z_{ij}'\mathbf {b}+K_{ij} \bigr) -
\operatorname{log} \bigl(Y_{ij}^{(u,v)}! \bigr) \bigr\}
\nonumber
\end{eqnarray}
based on the GLMM model. Since the $(u,v)$ pairs are independent of
each other, the complete likelihood, or the likelihood of $\mathbf
{Y}_{\mathrm{mis}}$, is
%
\begin{equation}
\label{completelikelihood} \ell_{c}(\mathbf{Y}_{\mathrm{mis}} | \utheta,
\mathbf{b}) = \operatorname{exp} \Biggl(-\sum_{u=0}^{1}
\sum_{v=0}^{1} h_{uv}(
\mathbf{Y}_{uv},\utheta,\mathbf {b}) \Biggr).
\end{equation}
The observed likelihood, or the likelihood of $\mathbf{Y}_{\mathrm{obs}}$, is the
marginal of $\ell_{c}$ summing over $\mathbf{Y}_{\mathrm{mis}}\in\mathcal{M}$.
Thus, we have
%
\begin{eqnarray}
\label{likelihoodobs} \ell_{\mathrm{obs}}(\mathbf{Y}_{\mathrm{obs}} | \utheta,
\mathbf{b}) &=& \sum_{\mathbf
{Y}_{\mathrm{mis}}\in\mathcal{M}} \ell_{c}(
\mathbf{Y}_{\mathrm{mis}} | \utheta,\mathbf{b}) = e^{ - h(\utheta,\mathbf{b})}
\end{eqnarray}
with
\begin{eqnarray*}
h(\utheta,\mathbf{b})& =&- \sum_{(i,j) \in\mathcal{I}} \bigl\{ \bigl(H_{ij}(\utheta)
+ z_{ij}'\mathbf{b} + K_{ij}
\bigr)Y_{ij} \\
&&\hspace*{41pt}{}- \operatorname {exp} \bigl(x_{ij}'(u,v)
\utheta+ z_{ij}'\mathbf{b}+K_{ij} \bigr) -
\operatorname {log} \bigl(Y_{ij}^{}! \bigr) \bigr\}
\end{eqnarray*}
and
%
\begin{equation}
\label{Hfunction} H_{ij}(\utheta) = \operatorname{log} \sum
_{u=0}^{1}\sum_{v=0}^{1}
\operatorname {exp} \bigl\{ x_{ij}'(u,v)\utheta \bigr\};
\end{equation}
the last equality in (\ref{likelihoodobs}) results from
simplifications based
on the well-known multinomial formula. Finally, marginalizing over
$\mathbf{b}$,
the marginal likelihood for $\uY_{\mathrm{obs}}$ given $\utau= (\utheta
',\sigma
^{2})'$, denoted by $\ell(\utau)$, becomes
%
\begin{equation}
\label{likelihood} \ell(\utau) = \int_{R^{F}} \ell_{\mathrm{obs}}(
\mathbf{Y}_{\mathrm{obs}} | \utheta, \mathbf{b})
\frac{e^{-{\mathbf{b}'\mathbf{b}}/{(2\sigma^{2})}}}{
(2\pi)^{F/2}\sigma^{F}} \,d\mathbf{b} =
\int_{R^{F}} \operatorname{exp} \bigl\{ -g(\utau,\mathbf{b})
\bigr\} \,d \mathbf{b},
\end{equation}
where
%
\begin{eqnarray}
\label{gdefinition} g(\utau,\mathbf{b}) &= &h(\utheta,\mathbf{b}) + h_{1}
\bigl(\sigma ^{2},\mathbf{b} \bigr),
\\
\label{h1definition} h_{1} \bigl(\sigma^{2},\mathbf{b} \bigr) &=&
\frac{1}{2\sigma^{2}}\mathbf {b}'\mathbf {b} + \frac{F}{2}
\operatorname{log} \sigma^{2} + \frac{F}{2}\operatorname {log}(2
\pi)
\end{eqnarray}
and the dependence of $\ell(\utau)$ on $\uY_{\mathrm{obs}}$ is suppressed for
convenience of the subsequent presentation.

\subsection{Laplace approximation of the likelihood and Bayesian inference}
\label{section:laplaceapproximationandbayesianinference}
The typical approach for obtaining inference on $\utau$ in a Bayesian
framework is to utilize the Gibbs sampler. The Gibbs sampler augments
the random effects vector $\mathbf{b}$ as additional parameters to be
estimated and, on convergence, gives samples from the posterior
distribution of $(\utau,\mathbf{b})$. This parameter augmentation step
avoids computing the integral in the likelihood (\ref{likelihood}).
However, in the case of fingerprint databases, $F$, the number of
fingers in a database, is typically large. As a result, parameter
augmentation schemes such as the Gibbs sampler take a considerably long
time to run until convergence and hinder any possibility of real time inference.

We avoid using any parameter augmentation scheme for the inference on
$\utau$. Our approach is to derive an approximation of the GLMM
likelihood $\ell(\utau)$ based on the Laplace expansion given by
%
\begin{equation}
\label{approximatelikelihood} \ell_{a}(\utau) = e^{-g(\utau, \hat{\mathbf{b}}(\utau) )} \biggl[
\operatorname{det} \biggl(\frac{1}{2\pi} \frac{\partial^{2} g(\utau, \hat
{\mathbf{b}}(\utau))}{\partial\utau^{2}} \biggr)
\biggr]^{-1/2},
\end{equation}
where $g(\utau,\mathbf{b})$ is the function defined in (\ref
{likelihood}), $\hat{\mathbf{b}}(\utau)$ is the maximum likelihood
estimate of $\mathbf{b}$ for fixed $\utau$, and ${\partial^{2}
g(\utau,
\hat{\mathbf{b}}(\utau))}/{\partial\utau^{2}}$ is the matrix of second
order partial derivatives with respect to $\mathbf{b}$ evaluated at
$\mathbf{b} = \hat{\mathbf{b}}(\utau)$, see \citet{SM95}. In the supplemental article
[\citet{supp}], we show that
%
\begin{equation}
\label{orderofapproximation} \operatorname{log} \bigl(\ell(\utau) \bigr) = \operatorname{log}
\bigl( \ell_{a}(\utau) \bigr) \biggl(1 + O \biggl(\frac{1}{F^{}}
\biggr) \biggr)
\end{equation}
as $F\rightarrow\infty$, meaning that the Laplace approximation
$\operatorname
{log}(\ell_{a}(\utau))$ is accurate up to order $O(1/F)$ as
$F\rightarrow\infty$. Equation (\ref{orderofapproximation}) justifies
the use of the Laplace-based approximate likelihood in place of (\ref
{likelihood}) when $F$ is large.

A further approximation to $\operatorname{log}(\ell_{a}(\utau))$ is obtained by
observing that
%
\begin{eqnarray}
\label{approximatelikelihoodsum} \operatorname{log} \bigl(\ell_{a}(\utau) \bigr) &=& -g
\bigl(\utau , \hat{\mathbf{b}}(\utau) \bigr) - \frac{1}{2} \operatorname{log
\, det} \biggl( \frac
{1}{2\pi} \frac{\partial^{2} g(\utau, \hat{\mathbf{b}}(\utau
))}{\partial\utau^{2}} \biggr)
\nonumber
\\[-8pt]
\\[-8pt]
\nonumber
&\equiv&(A) + (B)
\end{eqnarray}
is the sum of two terms: the first term $(A)$ involves a summation over
$F(F-1)L^{2}/2$ terms and, hence, is of that order as $F\rightarrow
\infty$. The order of the second term $(B)$ is $F \operatorname{log}(F-1 +
1/\sigma^{2})$ which follows from ${\partial^{2} g({\utau},\hat
{\mathbf
{b}}({\utau}))}/{\partial\utau^{2}} \sim(F-1 + 1/\sigma^{2})\uI_F$,
where $\uI_F$ is the $F \times F$ identity matrix. This is because each
diagonal entry of ${\partial^{2} g({\utau},\hat{\mathbf{b}}({\utau
}))}/{\partial\utau^{2}}$ involves a sum over $F-1$ terms together
with one term involving $1/\sigma^{2}$, whereas the off-diagonal
entries are each of order $1$. We retain the term $1/\sigma^{2}$ in the
case when $1/\sigma^{2}$ is of the same order as $F$ for small~$\sigma
^{2}$. Nevertheless, $F(F-1)L^{2}/2$ still dominates over $F \operatorname
{log}(F-1 + 1/\sigma^{2})$ for large $F$ and small $\sigma^{2}$ (where
$\sigma^{2} \sim F^{-1}$). This implies that $(A)$ dominates $(B)$ for
large $F$ and small $\sigma^{2}$ which motivates the further
approximation of $\operatorname{log}(\ell_{a}(\utau))$ by $(A)$. In Section~\ref{section:realdataanalysis}, we show that the above approximation is
valid based on the estimated values of $\sigma^{2}$ and choices of $F$
and $L$ for each database we worked with. The reader is also referred
to further details on the approximation presented in the supplemental
article [\citet{supp}].

We assume that the maximum likelihood estimate of $\utau$, $\hat
{\utau
}$, is available for the moment. Expanding $(A)$ in a Taylor's series
around $\hat{\utau}$, we get
%
\begin{equation}
\label{normalapproximatontau} g \bigl(\utau,\hat{\mathbf{b}}(\utau) \bigr)\approx g \bigl(\hat{
\utau},\hat {\mathbf {b}}(\hat{\utau}) \bigr) + \frac{1}{2}(\utau- \hat{
\utau})^{\prime} \biggl(\frac
{\partial^{2} g(\hat{\utau},\hat{\mathbf{b}}(\hat{\utau
}))}{\partial
\utau^{2}} \biggr) (\utau- \hat{\utau}),
\end{equation}
where ${\partial^{2} g(\hat{\utau},\hat{\mathbf{b}}(\hat{\utau
}))}/{\partial\utau^{2}}$ is the matrix ${\partial^{2} g({\utau
},\hat
{\mathbf{b}}({\utau}))}/{\partial\utau^{2}}$\vspace*{1pt} evaluated at $\utau=
\hat
{\utau}$. It is challenging to numerically evaluate $\hat{\utau}$ and
${\partial^{2} g(\hat{\utau},\hat{\mathbf{b}}(\hat{\utau
}))}/{\partial
\utau^{2}}$ in real time. This problem is addressed in greater detail
in Section~\ref{sectionmletau} subsequently.\vadjust{\goodbreak}

For inference in the Bayesian framework, we consider a prior $\pi_0$ on
$\utau$ of the product form: $\pi_0(\utau) = \pi_0(\utheta)\pi
_0(\sigma
^{2})$ with $\pi_0(\utheta) \propto1$ and $\pi_0(\sigma^{2})
\propto
(1/\sigma^{2}) I(\sigma^{2} > 0)$. These are standard noninformative
priors used on location and scale parameters in Bayesian literature.
Based on the likelihood and prior specifications, the exact and
approximate posteriors of $\utau$ (conditional on $\mathbf{Y}_{\mathrm{obs}}$) are
%
\begin{eqnarray}
\label{posterior} \pi(\utau| \mathbf{Y}_{\mathrm{obs}}) &=& \frac{\ell(\utau)\times\pi
_0(\utau)}{
\int_{\utau} \ell(\utau)\times\pi_0(\utau) \,d\utau}
\\
&\approx& \frac{\ell_{a}(\utau)\times\pi_0(\utau)}{
\int_{\utau} \ell_{a}(\utau)\times\pi_0(\utau) \,d\utau}
\\
\label{approximateposterior}&=&\pi_{a}(\utau| \mathbf{Y}_{\mathrm{obs}}),
\end{eqnarray}
say, based on equation (\ref{orderofapproximation}). The first
likelihood $\ell(\utau)$ is difficult to evaluate due to the integral
with respect to $\mathbf{b}$, whereas the second likelihood $\ell
_{a}(\utau)$ is easier to evaluate for each $\utau$ but difficult to
simulate from. However, based on equations (\ref
{approximatelikelihoodsum}) and (\ref{normalapproximatontau}), we
obtain samples of $\utau$ from $\tilde{\pi}(\utau)$, the multivariate
normal distribution with mean $\hat{\utau}$ and covariance matrix
${\partial^{2} g(\hat{\utau},\hat{\mathbf{b}}(\hat{\utau
}))}/{\partial
\utau^{2}}$. An importance sampling scheme is then used to convert
these realizations to posterior samples from $\pi_{a}(\utau| \mathbf
{Y}_{\mathrm{obs}} )$. More specifically, suppose $\utau_{h}^{*}$,
$h=1,2,\ldots,H$ are $H$ samples from $\tilde{\pi}$. Define the $H$ weights
$w_{h}^{*}$, $h=1,2,\ldots,H$ by
%
\begin{equation}
\label{weights} w_{h}^{*} = \frac{\pi_{a}(\utau_{h}^{*})}{D \tilde{\pi}(\utau_{h}^{*})},
\end{equation}
%
where $D = \sum_{h=1}^{H} {\pi_{a}(\utau_{h}^{*})}/{\tilde{\pi
}(\utau
_{h}^{*})}$ is the normalizing constant so that $\sum_{h=1}^{H}
w_{h}^{*}=1$. To obtain a sample from $\pi_a(\utau| \mathbf
{Y}_{\mathrm{obs}} )$ in (\ref{approximateposterior}), we resample from the
collection $\utau_{h}^{*}$ with weights $w_{h}^{*}$ for $h=1,2,\ldots,H$. This procedure is repeated $R$ times to obtain $R$ samples from
$\pi_a(\utau| \mathbf{Y}_{\mathrm{obs}})$. Numerical analysis showed that the
weights $w_{h}^{*}$ are uniformly distributed with value around $1/H$,
which indicates the effectiveness of approximating the target posterior
density $\pi_{a}$ using the Gaussian approximation~$\tilde{\pi}$.

\subsection{Obtaining \texorpdfstring{$\hat{\tau}$}{hat tau} using EM algorithm}
\label{sectionmletau}

The previous sections assume the availability of the maximum
likelihood estimate $\hat{\utau}$ of $\utau$ which we now describe how
to obtain. The estimator $\hat{\utau}$ is obtained based on the
maximizing function $-g(\utau,\hat{\mathbf{b}}(\utau))$. From equations
(\ref{orderofapproximation}) and~(\ref{approximatelikelihoodsum}), it
is clear that $\hat{\utau}$ also approximately maximizes the
log-likelihood function $\operatorname{log}(\ell_{a}(\utau))$ and,
subsequently, $\operatorname{log}(\ell(\utau))$, for large $F$. Ignoring
constant terms, we note that
%
\begin{eqnarray}
-g \bigl(\utau,\hat{\mathbf{b}}(\utau) \bigr) &=& -h \bigl(\utheta,\hat{\mathbf
{b}}(\utau) \bigr) - h_{1} \bigl(\utau,\hat{\mathbf{b}}(\utau) \bigr)
\\
\label{eqnEM} &=& \operatorname{log} \biggl(\sum_{\mathbf{Y}_{\mathrm{mis}}\in
\mathcal
{M}}
\ell_{c} \bigl(\mathbf{Y}_{\mathrm{mis}} | \utheta,\hat{\mathbf{b}}(
\utau ) \bigr) \biggr) - h_{1} \bigl(\sigma^{2},\hat{
\mathbf{b}}(\utau) \bigr),
\end{eqnarray}
where the function $h(\utheta,\mathbf{b})$ and $\ell_{c}$ are as
defined in
(\ref{likelihoodobs}) and
%
\begin{equation}
\label{h1} h_{1} \bigl(\sigma^{2},\mathbf{b} \bigr) =
\frac{1}{2\sigma^{2}}\mathbf {b}'\mathbf {b} + \frac{F}{2}
\operatorname{log} \sigma^{2}.\vadjust{\goodbreak}
\end{equation}
Equation (\ref{eqnEM}) sets the stage for an EM algorithm to be used:
Start with an initial estimate $\utau=\utau_{0}=(\utheta_{0},\sigma
_{0}^{2})$ and obtain $\utau= \utau_{k} = (\utheta_{k},\sigma
^{2}_{k})$ at the end of the $k$th step. At step $(k+1)$, the $E$-step
is carried out by noting that the conditional distribution of $ (
Y_{ij}^{(u,v)},u=0,1, v=0,1  )$ is multinomial with total number
of trials $Y_{ij}$ and multinomial probabilities
\[
p_{ij,k}^{(u,v)} = \frac{e^{x_{ij}(u,v)'\utheta_{k}}}{\sum_{u=0}^{1}\sum_{v=0}^{1} e^{x_{ij}(u,v)'\utheta_{k}}}
\]
independently for each $(i,j) \in\mathcal{I}$. Subsequently, we plug
in the expected value of each $Y_{ij}^{(u,v)}$, $Y_{ij,k}^{(u,v)}
\equiv Y_{ij} p_{ij,k}^{(u,v)}$ in place of $Y_{ij}^{(u,v)}$ in (\ref
{huv}) and (\ref{completelikelihood}). The $M$-step now entails
maximizing the objective function $-g_{c}(\utau,\hat{\mathbf
{b}}(\utau
))$ with respect to $\utau$, where
%
\begin{equation}
\label{objectivefunctionEM} g_{c}(\utau,\mathbf{b}) = h_{c}(\utheta,
\mathbf{b}) + h_{1} \bigl(\sigma ^{2},\mathbf{b} \bigr)
\end{equation}
and $h_{c}(\utheta,\mathbf{b})$ is given by
%
\begin{eqnarray}
\label{hc}&& h_{c}(\utheta,\mathbf{b}) = \sum
_{u=0}^{1}\sum_{v=0}^{1}
\sum_{(i,j)\in
\mathcal{I}} \bigl\{ \bigl( x_{ij}'(u,v)
\utheta+ z_{ij}'\mathbf{b} +K_{ij} \bigr)
Y_{ij,k}^{(u,v)}
\nonumber
\\[-8pt]
\\[-8pt]
\nonumber
&&\hspace*{114pt}{} + \operatorname{exp} \bigl(x_{ij}'(u,v)
\utheta+ z_{ij}'\mathbf {b}+K_{ij} \bigr) \bigr
\}.
\end{eqnarray}
This maximization yields $\utau= \utau_{k+1}$. Proceeding with
$k=1,2,\ldots $ gives $\utau= \hat{\utau}$ at convergence.

The $M$-step, or maximization of $-g_{c}(\utau,\hat{\mathbf
{b}}(\utau
))$ with respect to $\utau$, is carried out using the Newton--Raphson
procedure involving the first and second order partial derivatives of
$g_{c}$ with respect to $\utau$. At step $(k+1)$ of the EM algorithm,
we start with the initial value $\utau\equiv\utau_{k+1}^{0} \equiv
\utau_{k}$ as defined above. At step $(l+1)$, the current value $\utau
_{k+1}^{l+1}$ is obtained from $\utau_{k+1}^{l}$ using the equation
%
\begin{equation}
\label{NRiteration} \utau_{k+1}^{l+1} = \utau_{k+1}^{l}
- \biggl[\frac{\partial^{2}
g_{c}}{\partial\utau^{2}} \bigl(\utau_{k+1}^{l},\hat{
\mathbf{b}} \bigl(\utau _{k+1}^{l} \bigr) \bigr)
\biggr]^{-1} \frac{\partial g_{c}}{ \partial\utau
} \bigl(\utau _{k+1}^{l},
\hat{\mathbf{b}} \bigl(\utau_{k+1}^{l} \bigr) \bigr),
\end{equation}
where $\frac{\partial^{2} g_{c}}{\partial\utau^{2}}(\utau
_{k+1}^{l},\hat{\mathbf{b}}(\utau_{k+1}^{l}))$ and $\frac{\partial
g_{c}}{ \partial\utau}(\utau_{k+1}^{l},\hat{\mathbf{b}}(\utau
_{k+1}^{l}))$ are, respectively, the first and second order partial
derivatives of $g_{c}$ evaluated at $\utau=\utau_{k+1}^{l}$. The
explicit expressions for the first and second order partial derivatives
of $g_{c}(\utau,\hat{\mathbf{b}}(\utau))$ with respect to $\utau$ are
given in the supplemental article \citet{supp}. These
expressions, in turn, involve the first and second order partial
derivatives of $\hat{\mathbf{b}}(\utau)$ with respect to $\utau$. Since
no analytical form of $\hat{\mathbf{b}}(\utau)$ as a function of
$\utau
$ is available, one has to resort to numerical methods to estimate
$\hat
{\mathbf{b}}(\utau)$, its first and second order partial derivatives at
each iterative step of $k \ge0$ and $l\ge0$. A fast and effective way
of obtaining these numerical estimates is outlined in the supplemental
article \citet{supp} for the interested reader. On
convergence at $\utau=\hat{\utau}$, one obtains the numerical estimate
of the matrix ${\partial^{2} g(\hat{\utau},\hat{\mathbf{b}}(\hat
{\utau
}))}/{\partial\utau^{2}}$ by a similar method. The reader is referred
to the supplemental article \citet{supp} for details.

\section{Bayesian inference for the PRC}
\label{section:PRC}

Suppose $w$ minutia matches are observed for a fingerprint pair with
total number of detected minutiae $m_1$ and $m_2$, and with quality
measures $Q_{1}$ and $Q_2$ (assume continuous for the moment). Due to
varying image quality, not all of the $w$ matches correspond to matches
between genuine (true) minutiae. The model developed in Section~\ref{section:glmm} gives the number of true minutia matches, $Y^{(0,0)}$,
to be binomially distributed with parameters $w$ and $p_{00} \equiv
e^{\eta^{(0,0)}}/\sum_{u=0}^{1}\sum_{v=0}^{1} e^{\eta^{(u,v)}}$ for
the number of trials and success probability, respectively, where $\eta
^{(u,v)}$ are as defined in (\ref{sim1})--(\ref{sim4}). The binomial
distribution for $Y^{(0,0)}$ results by observing that the conditional
distribution of independent Poisson\vspace*{1pt} random variables $( Y^{(u,v)},
u,v=\{0,1\} )$ given their sum $Y= \sum_{u=0}^{1}\sum_{v=0}^{1}
Y^{(u,v)}=w$ is multinomial with total number of trials $w$ and
probabilities $p_{uv}\equiv e^{\eta^{(u,v)}}/\sum_{u=0}^{1}\sum_{v=0}^{1} e^{\eta^{(u,v)}}$ summing to one. It follows that the
marginal distribution of each $Y^{(u,v)}$ is binomial for each $(u,v) =
(0,0), (0,1), (1,0)$ and $(1,1)$. Assuming $Y^{(0,0)}$ is known, the
PRC corresponding to $Y^{(0,0)}$ matches is given by
%
\begin{equation}
\label{PRCexpression2} \operatorname{PRC}^{*} \bigl(Y^{(0,0)} | b_1,b_2,m_1,m_2,
\utau \bigr) = P \bigl(\mathcal{S} \ge Y^{(0,0)} \bigr),
\end{equation}
where $\mathcal{S}^{} \sim\operatorname{Poisson}(m_1 m_2 e^{\eta})$ and
$\eta= 2\beta_{0} + b_{1} + b_{2}$ with $b_{1}$ and $b_{2}$
distributed as independent $N(0,\sigma^{2})$ random variables. The
notation of $\operatorname{PRC}^{*}(\cdot| \cdots)$ in (\ref{PRCexpression2})
emphasizes its dependence on the GLMM parameters $\utau$, the
unobserved matches between genuine minutiae, $Y^{(0,0)}$, and the
random effects parameters $b_1$ and $b_2$. Since $(Y^{(0,0)},b_1,b_2)$
are unknown, the unconditional PRC is obtained by integrating out all
unknown random parameters, that is,
%
\begin{equation}
\label{PRCun} \operatorname{PRC}(w | m_1,m_2,\utau) =
E_{b_1,b_2,Y^{(0,0)}} \bigl[\operatorname{PRC}^{*} \bigl(Y^{(0,0)} |
b_1,b_2,m_1,m_2,\utau \bigr)
\bigr],
\end{equation}
where the expectation is taken over the joint distribution of
$(Y^{(0,0)},b_{1},b_{2})$ given~$\utau$. The random variables $b_1$ and
$b_2$ are independent from each other based on our modeling
assumptions, but note further that $Y^{(0,0)}$ is independent of
$(b_1,b_2)$. This is because the $p_{00}$ parameter of the binomial
distribution of $Y^{(0,0)}$, conditional on $b_1$ and $b_2$, is
independent of $b_1$ and $b_2$ (they cancel out from the numerator and
denominator expressions of $p_{00}$). Thus, for given $\utau$, it is
easy to sample from the joint distribution $(Y^{(0,0)},b_{1},b_{2})$
and estimate the expectation using a Monte Carlo sum. To obtain
inference for the PRC, assume $M$ samples from the posterior $\pi
_{a}(\utau| \uY_{\mathrm{obs}})$, $\utau_{1},\utau_{2},\ldots,\utau_{M}$, are
available. For $r=1,2,\ldots,R$, we obtain the $r$th sample of the PRC,
$\operatorname{PRC}_{r} \equiv \operatorname{PRC}(w | m_1,m_2,\utau_{r})$, by plugging $\utau=\utau
_{r}$ in~(\ref{PRCun}). These $R$ posterior samples are then used to
obtain the PRC posterior mean, standard deviation and the $100(1-\alpha
)\%$ credible intervals (mean $\pm z_{1-\alpha/2} $ sd) in Section~\ref{section:realdataanalysis}. Note that $\operatorname{PRC}(w | m_1,m_2,\utau)$ in
(\ref{PRCun}) should be the same value whether we use the combination
$(Q_1,Q_{2})$ or $(Q_2,Q_{1})$. Although this symmetry can be
established mathematically, small deviations away from symmetry arise
due to sampling error when (i) approximating the expected value in
(\ref
{PRCun}) using Monte Carlo, and (ii) using different random samples
from the posterior of $\utau$ for the two combinations $(Q_1,Q_2)$ and
$(Q_2,Q_1)$. Thus, a common estimate is reported in Section~\ref{section:realdataanalysis} which is the average of the estimates
obtained for the two combinations $(Q_1,Q_2)$ and $(Q_2,Q_1)$.

\section{Data analysis}
\label{section:realdataanalysis}

The publicly available databases provided by the Fingerprint
Verification Competitions [FVCs, \citet{finger:FVC2002} and
\citeauthor{FVC2006}] are considered
here. Specifically, subsets DB1 and DB2 of FVC2002 and subset DB3 of
FVC2006 are used. The FVC2002 DB1 database consists of fingerprint images
of $F=100$ different fingers and $L=8$ impressions per finger
obtained using the optical sensor ``TouchView II'' by Identix with
image size $388 \times374$ and resolution $500$ dots per inch (dpi).
The FVC2002 DB2 database consists of $L=8$ impressions from
$F=100$ fingers collected from an optical sensor ``FX2000'' by
Biometrika with image size $296 \times560$ and resolution $569$ dpi.
Fingerprint images in the DB1 and DB2 subsets were collected under
exaggerated distortions but were of good quality in general. The
FVC2006 database is comprised of four subsets, DB1 through DB4, each
consisting of $F=150$ fingers with $L=12$
impressions per finger. Fingerprint image data collection in FVC2006
was carried out using a thermal sweeping sensor with resulting image
size $400\times500$ and resolution $500$ dpi. 
Examples of images from the FVC2002 and FVC2006 databases are shown in
Figure~\ref{fig:variousimages}. Note the variability in the image
acquisition process due to using different sensors. Our methodology is
applied to each of the three subsets with two quality measures: (i) the
NFIQ categorical quality measure and (ii) the continuous quality
measure described in Section~\ref{empiricalfindings}. Results of the
parameter inference methodology for components of $\utau$ are as given
in Tables~\ref{table:parameterestimation:DB1}, \ref
{table:parameterestimation:DB2} and \ref{table:parameterestimation:DB3}
based on $R=200$ samples from the posterior distribution $\pi_{a}(
\utau| \uY_{\mathrm{obs}} )$.

\begin{table}
\tabcolsep=0pt
\caption{Posterior means ($M$),
standard deviations ($\mathit{SD}$) and $99.9\%$ credible intervals ($\mathit{CI}$) for
components of $\utau$ based on FVC2002 DB1}
\label{table:parameterestimation:DB1}
\begin{tabular*}{\textwidth}{@{\extracolsep{\fill}}lccclccc@{}}
\hline
&\multicolumn{3}{c}{\textbf{Continuous,} $\bolds{Q_{\mathrm{con}}}$} &
&\multicolumn{3}{c@{}}{\textbf{Categorical,} $\bolds{Q_{\mathrm{cat}}}$}\\ [-6pt]
&\multicolumn{3}{c}{\hrulefill} &
&\multicolumn{3}{c@{}}{\hrulefill}\\
$\bolds{\utau}$ & \textbf{M} & \textbf{SD} & \textbf{CI} & $\bolds{\utau}$ & \textbf{M} & \textbf{SD} & \textbf{CI}\\
\hline
$\theta_0$ & $-1.2801$ & $0.0061$ & $[-1.3003, -1.2599]$ & $\theta_0$ &
$-3.4857$ & $0.0049$ & $[-3.5018,-3.4697]$\\
$\theta_1$ & $-5.8520$ & $0.0137$ & $[-5.8969,-5.8070]$ & $\theta_1$ &
$-0.7429$ & $0.0013$ & $[-0.7472,-0.7386]$ \\
$-$ & $-$ & $-$ & $-$ & $\theta_2$ & $-1.6144$ & $0.0001$ &
$[-1.6145,-1.6143]$\\
$\beta_0$ & $-2.9047$ & $0.0065$ & $[-2.9259,-2.8835]$ & $\beta_0$ &
$-2.7297$ & $0.0064$ & $[-2.7507,-2.7087]$\\
$\operatorname{log}(\sigma^{2})$ & $-4.9518$ & $0.1460$ &
$[-5.4321,-4.4716]$ & $\operatorname{log}(\sigma^{2})$ & $-4.9537$ & $0.1231$
& $[-5.3588,-4.5486]$ \\
\hline
\end{tabular*}
\end{table}
%
\begin{table}
\tabcolsep=0pt
\caption{Posterior means ($M$),
standard deviations ($\mathit{SD}$) and $99.9\%$ credible intervals ($\mathit{CI}$) for
components of $\utau$ based on FVC2002 DB2}
\label{table:parameterestimation:DB2}
\begin{tabular*}{\textwidth}{@{\extracolsep{\fill}}lccclccc@{}}
\hline
&\multicolumn{3}{c}{\textbf{Continuous,} $\bolds{Q_{\mathrm{con}}}$} &
&\multicolumn{3}{c@{}}{\textbf{Categorical,} $\bolds{Q_{\mathrm{cat}}}$}\\ [-6pt]
&\multicolumn{3}{c}{\hrulefill} &
&\multicolumn{3}{c@{}}{\hrulefill}\\
$\bolds{\utau}$ & \textbf{M} & \textbf{SD} & \textbf{CI} & $\bolds{\utau}$ & \textbf{M} & \textbf{SD} & \textbf{CI}\\
\hline
$\theta_0$ & $-0.6346$ & $0.0020$ & $[-0.6413,-0.6280]$ & $\theta_0$ &
$-2.9255$ & $0.0019$ & $[-2.9319,-2.9191]$\\
$\theta_1$ & $-9.2982$ & $0.0024$ & $[-9.3062,-9.2901]$ & $\theta_1$ &
$-1.1496$ & $0.0011$ & $[-1.1532,-1.1460]$ \\
$-$ & $-$ & $-$ & $-$ & $\theta_2$ & $-0.9676$ & $0.0007$ &
$[-0.9698,-0.9653]$\\
$-$ & $-$ & $-$ & $-$ & $\theta_3$ & $-4.2827$ & $0.0007$ &
$[-4.2850,-4.2804]$\\
$\beta_0$ & $-2.8721$ & $0.0015$ & $[-2.8769,-2.8672]$ & $\beta_0$ &
$-2.8810$ & $0.0001$ & $[-2.8811,-2.8810]$\\
$\operatorname{log}(\sigma^{2})$ & $-4.2974$ & $0.0185$ &
$[-4.3581,-4.2366]$ & $\operatorname{log}(\sigma^{2})$ & $-4.7817$ & $0.0132$
& $[-4.8252,-4.7382]$ \\
\hline
\end{tabular*}
\end{table}

%
\begin{table}
\tabcolsep=0pt
\caption{Posterior means ($M$),
standard deviations ($\mathit{SD}$) and $99.9\%$ credible intervals ($\mathit{CI}$) for
components of $\utau$ based on FVC2006 DB3}
\label{table:parameterestimation:DB3}
\begin{tabular*}{\textwidth}{@{\extracolsep{\fill}}lccclccc@{}}
\hline
&\multicolumn{3}{c}{\textbf{Continuous,} $\bolds{Q_{\mathrm{con}}}$} &
&\multicolumn{3}{c@{}}{\textbf{Categorical,} $\bolds{Q_{\mathrm{cat}}}$}\\ [-6pt]
&\multicolumn{3}{c}{\hrulefill} &
&\multicolumn{3}{c@{}}{\hrulefill}\\
$\bolds{\utau}$ & \textbf{M} & \textbf{SD} & \textbf{CI} & $\bolds{\utau}$ & \textbf{M} & \textbf{SD} & \textbf{CI}\\
\hline
$\theta_0$ & $-6.1444$ & $0.0025$ & $[-6.1526,-6.1362]$ & $\theta_0$ &
$-5.4501$ & $0.0014$ & $[-5.4547,-5.4456]$\\
$\theta_1$ & $-3.6746$ & $0.0042$ & $[-3.6885,-3.6607]$ & $\theta_1$ &
$-0.2537$ & $0.0009$ & $[-0.2565,-0.2509]$ \\
$-$ & $-$ & $-$ & $-$ & $\theta_2$ & $-1.1425$ & $0.0011$ &
$[-1.1462,-1.1389]$\\
$-$ & $-$ & $-$ & $-$ & $\theta_3$ & $-2.3753$ & $0.0032$ &
$[-2.3859,-2.3647]$\\
$-$ & $-$ & $-$ & $-$ & $\theta_4$ & $-2.9186$ & $0.0029$ &
$[-2.9281,-2.9091]$\\
$\beta_0$ & $-3.1624$ & $0.0011$ & $[-3.1660,-3.1587]$ & $\beta_0$ &
$-3.1734$ & $0.0009$ & $[-3.1763,-3.1705]$\\
$\operatorname{log}(\sigma^{2})$ & $-3.9248$ & $0.0127$ &
$[-3.9666,-3.8831]$ & $\operatorname{log}(\sigma^{2})$ & $-4.2314$ & $0.0065$
& $[-4.2527,-4.2101]$ \\
\hline
\end{tabular*}
\end{table}

%
%


\begin{table}
\caption{Inference on
$\operatorname{PRC}(12 | 38,38)$ based on $Q_{\mathrm{cat}}$ and FVC2002 DB1: $M$ and $\mathit{CI}$
are, respectively, the posterior means and $99.9\%$ credible intervals}
\label{table:real:categorical:PRC:2002:db1}
\begin{tabular*}{\textwidth}{@{\extracolsep{\fill}}lcccccc@{}}
\hline
 & \multicolumn{2}{c}{$\mathbf{1}$} & \multicolumn
{2}{c}{$\mathbf{2}$} & \multicolumn{2}{c@{}}{$\mathbf{3}$} \\[-6pt]
 & \multicolumn{2}{c}{\hrulefill} & \multicolumn
{2}{c}{\hrulefill} & \multicolumn{2}{c@{}}{\hrulefill} \\
$\bolds{Q_1\setminus Q_2}$& $\mathbf{M}$ & $\mathbf{CI}$ & $\mathbf{M}$ & $\mathbf{CI}$ & $\mathbf{M}$ & $\mathbf{CI}$ \\
\hline
$1$ & $0.4082$ & $[0.3759$, & $0.2653$ & $[0.2424$, & $0.1541$ &
$[0.1325$,\\
& & $0.4404 ]$ & & $0.2882]$ & & $0.1758]$\\
$2$ & $0.2653$ & $[0.2424$, & $0.1383$ & $[0.1113$, & $0.0565$ &
$[0.0451$,\\
& & $0.2882]$ & & $0.1654]$ & & $0.0679]$ \\
$3$ & $0.1541$ & $[0.1325$, & $0.0565$ & $[0.0451$, & $0.0130$ &
$[0.0095$,\\
& & $0.1758]$ & & $0.0679]$ & & $0.0164]$\\
\hline
\end{tabular*}
\end{table}

\begin{table}[b]
\caption{Inference on
$\operatorname{PRC}(12 | 40,40)$ based on $Q_{\mathrm{cat}}$ and FVC2002 DB2: $M$ and $\mathit{CI}$
are, respectively, the posterior means and $99.9\%$ credible intervals}
\label{table:real:categorical:PRC:2002:db2}
\begin{tabular*}{\textwidth}{@{\extracolsep{\fill}}lcccccccc@{}}
\hline
 & \multicolumn{2}{c}{$\mathbf{1}$} & \multicolumn
{2}{c}{$\mathbf{2}$} & \multicolumn{2}{c}{$\mathbf{3}$} &\multicolumn{2}{c@{}}{$\mathbf{4}$} \\[-6pt]
 & \multicolumn{2}{c}{\hrulefill} & \multicolumn{2}{c}{\hrulefill} & \multicolumn{2}{c}{\hrulefill}& \multicolumn{2}{c@{}}{\hrulefill} \\
$\bolds{Q_1\setminus Q_2}$& $\mathbf{M}$ & $\mathbf{CI}$ & $\mathbf{M}$ & $\mathbf{CI}$ & $\mathbf{M}$ & $\mathbf{CI}$ &$\mathbf{M}$ & $\mathbf{CI}$\\
\hline
$1$ & $0.7958$ & $[0.7873$, & $0.5858$ & $[0.5733$, & $0.4734$ &
$[0.4605$, & $0.3881$ & $[0.3750$,\\
& & $0.8044]$ & & $0.5983]$ & & $0.4862]$ & & $0.4013]$\\
$2$ & $0.5858$ & $[0.5733$, & $0.2725$ & $[0.2609$, & $0.1556$ &
$[0.1469$, & $0.0902$ & $[0.0838$,\\
& & $0.5983]$ & & $0.2841]$ & & $0.1643]$ & & $0.0967]$\\
$3$ & $0.4734$ & $[0.4605$, & $0.1556$ & $[0.1469$, & $0.0667$ &
$[0.0608$, & $0.0275$ & $[0.0245$,\\
& & $0.4862]$ & & $0.1643]$ & & $0.0726]$ & & $0.0306]$ \\
$4$ & $0.3881$ & $[0.3750$, & $0.0902$ & $[0.0838$, & $0.0275$ &
$[0.0245$, & $0.0073$ & $[0.0062$,\\
& & $0.4013]$ & & $0.0967]$ & & $0.0306]$ & & $0.0085]$\\
\hline
\end{tabular*}
\end{table}

\begin{table}
\caption{Inference on
$\operatorname{PRC}(12 | 38,38)$ based on $Q_{\mathrm{con}}$ and FVC2002 DB1: $M$ and $\mathit{CI}$
are, respectively, the posterior means and $99.9\%$ credible intervals}
\label{table:real:continuous:PRC:2002:db1}
\begin{tabular*}{\textwidth}{@{\extracolsep{\fill}}lcccccc@{}}
\hline
 & \multicolumn{2}{c}{$\bolds{0.3}$} & \multicolumn
{2}{c}{$\bolds{0.4}$} & \multicolumn{2}{c@{}}{$\bolds{0.5}$} \\[-6pt]
 & \multicolumn{2}{c}{\hrulefill} & \multicolumn
{2}{c}{\hrulefill} & \multicolumn{2}{c@{}}{\hrulefill} \\
$\bolds{Q_1\setminus Q_2}$& $\mathbf{M}$ & $\mathbf{CI}$ & $\mathbf{M}$ & $\mathbf{CI}$ & $\mathbf{M}$ & $\mathbf{CI}$ \\
\hline
$0.3$ & $0.5358$ & $[0.5126$, & $0.3908$ & $[0.3671$, & $0.2868$ &
$[0.2650$,\\
& & $0.5591]$ & & $0.4146]$ & & $0.3087]$\\
$0.4$ & $0.3908$ & $[0.3671$, & $0.2383$ & $[0.2174$, & $0.1455$ &
$[0.1280$, \\
& & $0.4146]$ & & $0.2592]$ & & $0.1631]$ \\
$0.5$ & $0.2868$ & $[0.2650$, & $0.1455$ & $[0.1280$, & $0.0724$ &
$[0.0610$,\\
& & $0.3087]$ & & $0.1631]$ & & $0.0838]$\\
\hline
\end{tabular*}
\end{table}

\textit{Inference for the PRC}: We report the PRC corresponding to $w=12$
matches using the method outlined in Section~\ref{section:PRC}. Note
that $w=12$ is used for illustrative purposes only; similar inference
results on the PRC can be obtained for any observed number of minutia
matches $w$. The values of $m_1$ and $m_2$, the total number of
extracted minutiae in the two prints, are fixed at mean values for each
database: These are $(m_1,m_2)=(38,38), (40,40)$ and $(84,84)$ for
FVC2002 DB1, FVC2002 DB2 and FVC2006 DB3, respectively. The mean
numbers for FVC2002 DB1 and DB2 are similar, and are much smaller
compared to FVC2006 DB3. Tables~\ref{table:real:categorical:PRC:2002:db1} and \ref
{table:real:categorical:PRC:2002:db2} give the results for
FVC2002 DB1 and DB2 based on the categorical quality measure, whereas Tables~\ref{table:real:continuous:PRC:2002:db1} and \ref
{table:real:continuous:PRC:2002:db2} give the results for the
continuous quality measure. Similar reports and conclusions are
obtained for FVC2006 DB3 as in FVC2002 DB1 and DB2, so the relevant
tables are not presented here but relegated to the Appendix.

The posterior mean estimates of the $\operatorname{PRC}$ are monotonically decreasing
functions of increasing quality measures (for both NFIQ and the
continuous quality measure), as should be expected. As quality becomes
better, erroneous decisions due to spurious minutia matches are
reduced. In turn, this component of uncertainty in the PRC evaluation
is also decreased. Thus, for high quality images, the PRC essentially
captures the inherent inter-finger variability in the population. The
results reported in Tables~\ref{table:real:categorical:PRC:2002:db1}--\ref
{table:real:continuous:PRC:2002:db2} show the importance of having good
quality images when a decision of a positive match has to be made. With
the requirement of $\operatorname{PRC} \approx0.05$, we see from the four tables that
this is achieved when both prints are of the best quality. The PRC
deteriorates quickly as the image quality degrades, as is indicated by
all the four tables, signifying the importance of very good quality
images for PRC assessment. For poor quality images, one has to be extra
cautious interpreting the extent of a fingerprint match, as the
uncertainty associated with a positive match is very large.

\begin{table}[b]
\caption{Inference on
$\operatorname{PRC}(12 | 40,40)$ based on $Q_{\mathrm{con}}$ and FVC2002 DB2: $M$ and $\mathit{CI}$
are, respectively, the posterior means and $99.9\%$ credible intervals}
\label{table:real:continuous:PRC:2002:db2}
\begin{tabular*}{\textwidth}{@{\extracolsep{\fill}}lcccccccc@{}}
\hline
 & \multicolumn{2}{c}{$\bolds{0.2}$} & \multicolumn
{2}{c}{$\bolds{0.4}$} & \multicolumn{2}{c}{$\bolds{0.5}$}
&\multicolumn{2}{c@{}}{$\bolds{0.6}$} \\[-6pt]
 & \multicolumn{2}{c}{\hrulefill} & \multicolumn{2}{c}{\hrulefill} & \multicolumn{2}{c}{\hrulefill}& \multicolumn{2}{c@{}}{\hrulefill} \\
$\bolds{Q_1\setminus Q_2}$& $\mathbf{M}$ & $\mathbf{CI}$ & $\mathbf{M}$ & $\mathbf{CI}$ & $\mathbf{M}$ & $\mathbf{CI}$&$\mathbf{M}$ & $\mathbf{CI}$ \\
\hline
$0.2$ & $0.9152$ & $[0.9087$, & $0.6967$ & $[0.6811$, & $0.6292$ &
$[0.6134$, & $0.5965$ & $[0.5797$, \\
& & $0.9217]$ & & $0.7123]$ & & $0.6451]$ & & $0.6134]$\\
$0.4$ & $0.6967$ & $[0.6811$, & $0.1956$ & $[0.1816$, & $0.1158$ &
$[0.1051$, & $0.0873$ & $[0.0787$,\\
& & $0.7123]$ & & $0.2096]$ & & $0.1265]$ & & $0.0959]$\\
$0.5$ & $0.6292$ & $[0.6134$, & $0.1158$ & $[0.1051$, & $0.0559$ &
$[0.0489$, & $0.0371$ & $[0.0323$, \\
& & $0.6451]$ & & $0.1265]$ & & $0.0630]$ & & $0.0420]$ \\
$0.6$ & $0.5965$ & $[0.5797$, & $0.0873$ & $[0.0787$, & $0.0371$ &
$[0.0323$, & $0.0228$ & $[0.0193$,\\
& & $0.6134]$ & & $0.0959]$ & & $0.0420]$ & & $0.0264]$\\
\hline
\end{tabular*}
\end{table}

\begin{figure}

\includegraphics{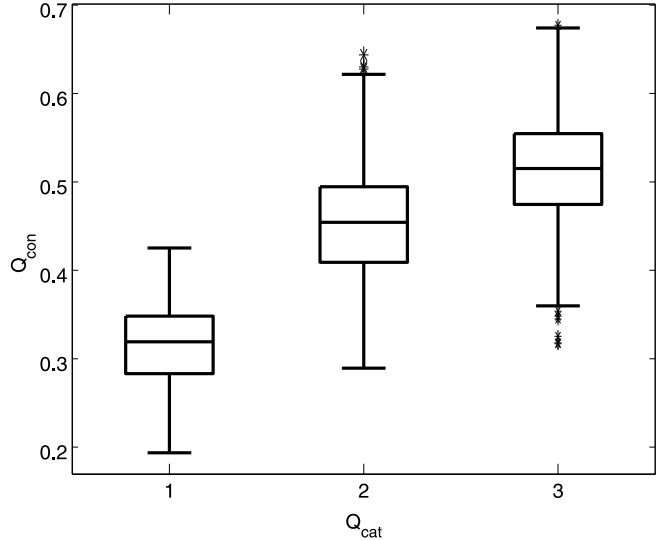}

\caption{Boxplots showing the distribution of
$Q_{\mathrm{con}}$ for each level of $Q_{\mathrm{cat}}=1, 2$ and $3$ for DB1.} \label{fig:boxplot}
\end{figure}

Tables~\ref{table:real:categorical:PRC:2002:db1}--\ref
{table:real:continuous:PRC:2002:db2} invite some comparisons. For DB1,
the average values of $Q_{\mathrm{con}}$ for each value of $Q_{\mathrm{cat}}=1,2,3$ are
$0.32,0.45,0.51$, respectively; for DB2, the average values of
$Q_{\mathrm{con}}$ for each value of $Q_{\mathrm{cat}}=1,2,3,4$ are $0.24$,
$0.38$, $0.53$ and $0.58$, respectively. The levels of $Q_{\mathrm{con}}$ chosen
in Tables~\ref{table:real:continuous:PRC:2002:db1} and \ref
{table:real:continuous:PRC:2002:db2} are taken to reflect these average
values. So, one would expect that the PRC results presented for each
database be consistent over $Q_{\mathrm{cat}}$ and $Q_{\mathrm{con}}$, especially when
the value of $\operatorname{PRC}$ is small, say, less than $0.05$, for example. This
is generally the case when we compare the entries of Tables~\ref{table:real:categorical:PRC:2002:db1} and \ref
{table:real:continuous:PRC:2002:db1} for DB1 and Tables~\ref{table:real:categorical:PRC:2002:db2} and \ref
{table:real:continuous:PRC:2002:db2} for DB2. Minor differences in the
PRCs can be attributed to the significant variability of $Q_{\mathrm{con}}$ for
each level of $Q_{\mathrm{cat}}$ in the three databases. While we would expect
$Q_{\mathrm{con}}$ values to be different corresponding to different levels of
$Q_{\mathrm{cat}}$, we do not find this to be the case empirically. There is
significant overlap between the $Q_{\mathrm{con}}$ levels for the different
$Q_{\mathrm{cat}}$ values. In the case when $Q_{\mathrm{cat}}=3$ for DB1, the range of
$Q_{\mathrm{con}}$ is from $0.32$ to $0.67$, whereas it is $0.30$ to $0.64$ for
$Q_{\mathrm{cat}}=2$; see Figure~\ref{fig:boxplot} for an idea of the range of
these values for $Q_{\mathrm{con}}$.

The PRCs corresponding to $Q_{\mathrm{con}}$ and $Q_{\mathrm{cat}}$ are much larger for
DB3 [see Tables~1 and 2 in the supplemental article \citet{supp}] 
compared to DB1 and DB2, because $(m_1,m_2)=(84,84)$, the mean values
for DB3, are much larger compared to the mean minutia numbers for DB1
and DB2. As a result, it is much more likely for random pairings to
occur in DB3 since more minutia are available for pairings.


\textit{Analysis and interpretation for the forensic practitioner}: We
illustrate how PRCs can be obtained for a pair of prints
under investigation in forensic practice. The pair in Figure~\ref{fig:impostor:match} has $w=7$ matches with $m_1=35$ and $m_2=49$, and
$Q_{\mathrm{cat}}=2$ and $3$ for the left and right images, respectively. Since
these images are from FVC2002 DB2, we base our inference using the
results from this database. Corresponding to $w=7$, the mean PRC is
$0.6395$ with $99.9\%$ credible interval $[0.6263, 0.6523]$. The mean
value implies that about $64\%$ of random pairings of impostor
fingerprints will have minutia matches of $7$ or more. Thus, there is
no evidence to suggest that the pair in Figure~\ref{fig:impostor:match}
is something other than a random match (namely, an impostor pair).

The forensic practitioner can also perform a ``what if'' analysis. If
the quality of the images was good with $Q_{\mathrm{cat}}$ taking value $4$ for
both images in the pair, then the PRC is reduced to $0.3112$, which
still indicates this is more likely a random pairing. With this
analysis, a practitioner can eventually conclude that the pairs in
question are very likely a random match even in the best case scenario
where the image qualities are perfect.

The GLMM procedure can also be utilized to provide a guideline for
choices of $w$ that will give desired PRC values for forensic
testimony, for example, $\operatorname{PRC}=0.01$ means that the error we make in
declaring a positive match when in fact it is not is at most $0.01$.
Table~\ref{table:designw} provides the smallest values for $w$ for
guaranteeing $\operatorname{PRC}(w | m_1,m_2) \le0.01$ for different $Q_{\mathrm{cat}}$
combinations when $m_1=35$ and $m_2=49$ based on DB2 with maximum
possible number of matches $w$ being ${\operatorname{min}}(m_1,m_2)=35$. Based on the
$(2,3)$ [or $(3,2)$] entry of the table, we see that the desired $w$ to
make a reliable decision for the above prints in question should be
$21$. Thus, $w=7$ is too low for making a reliable positive match
decision. The $*$ entries corresponding to the lowest quality
combinations indicate that even for the maximum value of $w=35$, the
PRC never goes below $0.01$. This again emphasizes that positive
identification decisions cannot and should not be made with very low
quality images. Corresponding results can be obtained for DB1 and DB3
similarly, and for $Q_{\mathrm{con}}$, and are therefore not presented here.

\begin{table}
\tablewidth=200pt
\caption{Smallest $w$ required for $\operatorname{PRC}(w |
35,49)\le0.01$ based on $Q_{\mathrm{cat}}$ and DB2. $*$ indicates no such $w$ exists}
\label{table:designw}
\begin{tabular*}{200pt}{@{\extracolsep{\fill}}lcccc@{}}
\hline
$\bolds{Q_1 / Q_2}$ & $\bolds{1}$ & $\bolds{2}$ & $\bolds{3}$ & $\bolds{4}$\\
\hline
$1$ & $*$ & $*$ & $33$ & $29$\\
$2$ & $*$ & $25$ & $21$ & $18$\\
$3$ & $33$ & $21$ & $17$ & $15$\\
$4$ & $29$ & $18$ & $15$ & $13$\\
\hline
\end{tabular*}
\end{table}

\section{Validation based on simulation}
\label{simulation:results}


We performed simulation and validation experiments for DB1 and DB2 with
$F=100$ fingers and $L=8$ impressions per finger. $Q_{\mathrm{cat}}$ and
$Q_{\mathrm{con}}$ levels were fixed as in the respective databases, but the
total number of minutia matches $Y_{ij}$ for each pair of prints
$(i,j)$ (see Section~\ref{section:glmm} for the GLMM model and related
notation) were simulated from the GLMM model with fixed and known
parameter values $\utau= (\utheta',\operatorname{log}(\sigma^{2}))'$. The
parameter values were fixed at the estimated values obtained based on
real data for DB1 and DB2, as given in Tables~\ref{table:parameterestimation:DB1} and \ref
{table:parameterestimation:DB2}. As noted previously, our inference
procedure yields posterior mean and standard deviation estimates as
well as $99.9\%$ credible intervals as mentioned in Section~\ref{section:BayesianInference}. The coverage probabilities of each
credible interval were calculated for all the true parameter values as
well as the true PRC values based on $50$ runs. We use the coverage
probabilities as a measure of how well the Laplace method approximates
the exact GLMM likelihood for large $F$. Based on simulations, we find
that the average coverage probabilities
for parameters and PRC values (averaged over all three databases) are
approximately $98\%$ and $98.5\%$, respectively. There is some
underestimation of the coverage probability but not grossly so,
indicating that the Laplace approximation to the GLMM likelihood is
good even for $F=100$.

\section{Summary and discussion}
\label{section:discussion}

To assess the extent of fingerprint individuality for different image
quality of prints, a GLMM framework and Bayesian inference procedure is
developed. Our inference scheme is able to provide a point estimate as
well as a credible interval for the PRC depending on the image quality
and the observed number of matches between a pair of prints. Numerical
reports of the PRC are obtained which can be used to validate forensic
testimony. Further, we also provide the smallest number of minutia
matches $w$ needed to keep the PRC around a prespecified (small)
number. These matching numbers serve as a guideline and safeguard
against falsely deciding a positive match when the quality of the
prints is unreliable. The best report of fingerprint individuality is
obtained when both or at least one is of very good quality and the
other is of good quality. No inference on individuality should be made
when either print is of moderate to poor quality as observed from our
experimental results; having poor quality latent prints, for example,
severely hampers reliable matching results.

As far as we know, previous work has not considered including image
quality in fingerprint individuality assessment in a quantitative
manner and in a formal statistical framework. Our
work can be used as a baseline in forensic applications for acquiring a
quantitative assessment of individuality given two prints in question.
Regardless of whether the prints are of poor or good quality, the
methodology outlined will output a PRC value. This PRC value could
serve as a baseline and as a first guide to what extent there is
evidence beyond a reasonable doubt. The subsection of Section~\ref{section:realdataanalysis} titled ``Analysis and inference for the
forensic practitioner'' highlights the kinds of analysis that may be
conducted based on the GLMM methodology. The GLMM method still does not
utilize other fingerprint features such as type of minutiae detected,
ridge lengths and the class of fingerprint, which could potentially
further decrease the PRC and increase the extent of individualization.
This will be addressed in a future work.

\textit{Intra-finger correlations and inter-finger variability}:  The
incorporation of the random effects $b_f$ is very important in any
statistical analysis involving different fingers and multiple
impressions per finger. The random effects $b_f$ serve to model both
intra-finger correlations as well as inter-finger variability. Models
that incorporate correlations due to multiple impressions of the same
finger give very different results compared to models that do not
account for this. For example, \citet{D10} show that the upper
and lower confidence bounds are misleadingly narrower if correlations,
when present, are not taken into account. Similarly, the upper and
lower credible bounds for the PRC will be affected if one does not
account for intra-finger correlations. The random effects $b_{f}$ also
serve another purpose. They model inter-finger variability as we move
from different fingers in the target population; the database is merely
a representative sample of the target population of fingers for which
the true population PRC is unknown and has to be estimated. In Zhu, Dass and Jain (\citeyear{ZDJ07}),
the PRC was calculated for each pair of print based on
minutia distributions on the respective fingers. However, this is not a
representative PRC for the \textit{entire} fingerprint database. Thus,
Zhu, Dass and Jain (\citeyear{ZDJ07}) considered clustering the minutia distributions for the
entire fingerprint database. A more formal approach for obtaining
inference for population level PRCs was addressed in \citet{D11}. The random effects $b_{f}$ in the present context represent
deviations in the target population, thus enabling inference for this
overall population PRC to be made.

\textit{Alignment of prints}: The alignment prior to matching a pair of
prints is not a separate issue but a function of the detected minutia.
Based on the detected minutia in a print, the alignment with the other
print is done by first aligning a pair of detected minutiae (one from
each print), then finding the optimal translation (via the paired
minutiae locations) and rotation (via the paired minutiae directions)
that exactly aligns the two. So, one can see that the alignment is also
a function of image quality: For poor quality images, more spurious
minutiae are detected and, hence, the alignment can be more random
(compared to the true alignment based on genuine minutia). The
randomness in the alignment also contributes to increasing the number
of minutia matches, $Y_{ij}$, for poor quality images. We do not model
the alignment separately because of the fact that its randomness, which
depends on spurious minutia, is captured in the observed $Y_{ij}$.

\textit{Fusion schemes}: The difference in the estimates of fingerprint
individuality for the three databases can be attributed to the
intrinsic nature of the images in these separate databases. For the
databases considered, the intrinsic variability arises due to the
different sensors used, the extent of distortion due to varying skin
elasticity and the composition of the subjects (manual workers versus
aged population and others). Where forensic application is concerned,
it is usually the practice to match a latent print to template prints
in a database. The sensor used to acquire the templates is known and,
therefore, it makes sense to report sensor-specific estimates of
fingerprint individuality. However, sensor-specific estimates should
not be too different from each other: A significant difference in the
reported PRCs (using sensor-specific models) indicates systematic bias
of that particular sensor toward or against random matching. As we
mentioned and observed earlier, we also find significant variability
(as well as overlap) in $Q_{\mathrm{con}}$ values corresponding to different
$Q_{\mathrm{cat}}$ levels. This motivates us to consider GLMMs and inference for
PRCs based on combining two or more quality covariates that measure
different aspects of image clarity. It is possible to arrive at a
single measure of fingerprint individuality by incorporating additional
random effects (for different fingerprint databases and sensors) and
fixed effects (for two or more quality measures) in the GLMM
formulation. We will investigate these fusion issues in our future research.

\section*{Acknowledgments}
\label{section:acknowledgment}
The authors thank the Editor, Associate Editor and referees for their
valuable suggestions in improving this paper.

\begin{supplement}[id=suppA]
\stitle{For the supplemental article ``A generalized mixed model
framework for assessing fingerprint individuality in presence of
varying image quality''}
\slink[doi]{10.1214/14-AOAS734SUPP} 
\sdatatype{.pdf}
\sfilename{aoas734\_supp.pdf}
\sdescription{The results quoted in the main text are proved in
Section~1
and the tables of PRC results for DB3 are in Section~2.}
\end{supplement}



\printaddresses

\begin{thebibliography}{15}

\bibitem[\protect\citeauthoryear{Dass}{2010}]{Dass10}
\begin{barticle}[auto:STB|2014/02/12|14:17:21]
\bauthor{\bsnm{Dass},~\bfnm{S.}\binits{S.}}
(\byear{2010}).
\btitle{Assessing fingerprint individuality in presence of noisy minutiae}.
\bjournal{IEEE Trans. of Information Forensics and Security}
\bvolume{5}
\bpages{62--70}.
\end{barticle}
\bptok{imsref}%
\endbibitem

\bibitem[\protect\citeauthoryear{Dass and Jain}{2006}]{D10}
\begin{barticle}[auto:STB|2014/02/12|14:17:21]
\bauthor{\bsnm{Dass},~\bfnm{S.}\binits{S.}} \AND
\bauthor{\bsnm{Jain},~\bfnm{A.~K.}\binits{A.~K.}}
(\byear{2006}).
\btitle{Validating a biometric authentication system: Sample size requirements}.
\bjournal{IEEE Transactions on Pattern Analysis and Machine Intelligence}
\bvolume{28}
\bpages{1902--1319}.
\end{barticle}
\bptok{imsref}%
\endbibitem

\bibitem[\protect\citeauthoryear{Dass and Li}{2009}]{D11}
\begin{barticle}[auto:STB|2014/02/12|14:17:21]
\bauthor{\bsnm{Dass},~\bfnm{S.}\binits{S.}} \AND
\bauthor{\bsnm{Li},~\bfnm{M.}\binits{M.}}
(\byear{2009}).
\btitle{Hierarchical mixture models for assessing
fingerprint individuality}.
\bjournal{Ann. Appl. Stat.}
\bvolume{3}
\bpages{1448--1466}.
\bid{mr={2752141}}
\end{barticle}
\bptok{imsref}%
\endbibitem

\bibitem[\protect\citeauthoryear{Dass, Lim and Maiti}{2014}]{supp}
\begin{bmisc}[author]
{\bauthor{\bsnm{Dass},~\binits{S. C.}},
\bauthor{\bsnm{Lim},~\binits{C.}} \AND
\bauthor{\bsnm{Maiti},~\binits{T.}}}
(\byear{2014}).
\bhowpublished{Supplement to ``A generalized mixed model framework for assessing
fingerprint individuality in presence of varying image
quality.''
DOI:\doiurl{10.1214/14-AOAS734SUPP}}.
\bptok{imsref}%
\end{bmisc}
\endbibitem

\bibitem[\protect\citeauthoryear{Daubert vs. Merrell Dow
Pharmaceuticals}{1995}]{Daubert}
\begin{bmisc}[auto:STB|2014/02/12|14:17:21]
\borganization{Daubert vs. Merrell Dow
Pharmaceuticals}
(\byear{1995}).
\bhowpublished{509 U.S. 579, 113 S. Ct. 2786, 125 L.Ed.2d 469}.
\end{bmisc}
\bptok{imsref}%
\endbibitem

\bibitem[\protect\citeauthoryear{FVC2006}{2006}]{FVC2006}
\begin{bmisc}[auto:STB|2014/02/12|14:17:21]
\borganization{FVC2006: Fingerprint Verification Competition}
(\byear{2006}).
\bhowpublished{\url{http://bias.csr.unibo.it/fvc2006/}.}
\end{bmisc}
\bptok{imsref}%
\endbibitem

\bibitem[\protect\citeauthoryear{Home Office Automatic Fingerprint Recognition System (HOAFRS), License 16-93-0026}{1993}]{LIC93}
\begin{bmisc}[auto:STB|2014/02/12|14:17:21]
\borganization{Home Office Automatic Fingerprint Recognition System (HOAFRS), License 16-93-0026}
(\byear{1993}).
\bhowpublished{\textit{Science and technology group}. Home Office, London}.
\end{bmisc}
\bptok{imsref}%
\endbibitem

\bibitem[\protect\citeauthoryear{Lehmann and Romano}{2005}]{Lehman}
\begin{bbook}[mr]
\bauthor{\bsnm{Lehmann},~\bfnm{E.~L.}\binits{E.~L.}} \AND
\bauthor{\bsnm{Romano},~\bfnm{Joseph~P.}\binits{J.~P.}}
(\byear{2005}).
\btitle{Testing Statistical Hypotheses},
\bedition{3rd} ed.
\bpublisher{Springer},
\blocation{New York}.
\bid{mr={2135927}}
\end{bbook}
\bptok{imsref}%
\endbibitem

\bibitem[\protect\citeauthoryear{Maio et~al.}{2002}]{finger:FVC2002}
\begin{bincollection}[auto:STB|2014/02/12|14:17:21]
\bauthor{\bsnm{Maio},~\bfnm{D.}\binits{D.}},
\bauthor{\bsnm{Maltoni},~\bfnm{D.}\binits{D.}},
\bauthor{\bsnm{Cappelli},~\bfnm{R.}\binits{R.}},
\bauthor{\bsnm{Wayman},~\bfnm{J.~L.}\binits{J.~L.}} \AND
\bauthor{\bsnm{Jain},~\bfnm{Anil~K.}\binits{A.~K.}}
(\byear{2002}).
\btitle{FVC2002: Fingerprint verification competition}.
In \bbooktitle{Proceedings of the International Conference on
Pattern Recognition (ICPR)}
\bpages{744--747}.
\bpublisher{IEEE Computer Society},
\blocation{Quebec, Canada}.
\end{bincollection}
\bptok{imsref}%
\endbibitem

\bibitem[\protect\citeauthoryear{National Academy of Sciences Committee on Identifying the Needs of the Forensic Science Community, National Research Council}{2009}]{NAS:2009}
\begin{bmisc}[auto:STB|2014/02/12|14:17:21]
\borganization{National Academy of Sciences Committee on Identifying the Needs of the Forensic Science Community, National Research Council}
(\byear{2009}).
\bhowpublished{Strengthening forensic science in the United States:
A path forward. National Academies Press}.
\end{bmisc}
\bptok{imsref}%
\endbibitem

\bibitem[\protect\citeauthoryear{Pankanti, Prabhakar and Jain}{2002}]{PPJ02}
\begin{barticle}[auto:STB|2014/02/12|14:17:21]
\bauthor{\bsnm{Pankanti},~\bfnm{S.}\binits{S.}},
\bauthor{\bsnm{Prabhakar},~\bfnm{S.}\binits{S.}} \AND
\bauthor{\bsnm{Jain},~\bfnm{A.~K.}\binits{A.~K.}}
(\byear{2002}).
\btitle{On the individuality of fingerprints}.
\bjournal{IEEE Transactions on Pattern Analysis and Machine Intelligence}
\bvolume{24}
\bpages{1010--1025}.
\end{barticle}
\bptok{imsref}%
\endbibitem

\bibitem[\protect\citeauthoryear{Shun and McCullagh}{1995}]{SM95}
\begin{barticle}[mr]
\bauthor{\bsnm{Shun},~\bfnm{Zhenming}\binits{Z.}} \AND
\bauthor{\bsnm{McCullagh},~\bfnm{Peter}\binits{P.}}
(\byear{1995}).
\btitle{Laplace approximation of high-dimensional integrals}.
\bjournal{J.~Roy. Statist. Soc. Ser. B}
\bvolume{57}
\bpages{749--760}.
\bid{issn={0035-9246}, mr={1354079}}
\end{barticle}
\bptok{imsref}%
\endbibitem

\bibitem[\protect\citeauthoryear{Tabassi, Wilson and Watson}{2004}]{TWW04}
\begin{bmisc}[auto:STB|2014/02/12|14:17:21]
\bauthor{\bsnm{Tabassi},~\bfnm{E.}\binits{E.}},
\bauthor{\bsnm{Wilson},~\bfnm{C.}\binits{C.}} \AND
\bauthor{\bsnm{Watson},~\bfnm{C.}\binits{C.}}
(\byear{2004}).
\bhowpublished{Fingerprint image quality.
Technical Report 7151.
Available at \url{http://fingerprint.nist.gov/NBIS}.}
\end{bmisc}
\bptok{imsref}%
\endbibitem

\bibitem[\protect\citeauthoryear{U.S. vs. Byron C. Mitchell}{1999}]{USvsByronCMitchell}
\begin{bmisc}[auto:STB|2014/02/12|14:17:21]
\borganization{U.S. vs. Byron C. Mitchell}
(\byear{1999}).
\bhowpublished{Criminal Action No. 96-407, U.S. District Court for the Eastern District of Pennsylvania}.
\end{bmisc}
\bptok{imsref}%
\endbibitem

\bibitem[\protect\citeauthoryear{Zhu, Dass and Jain}{2007}]{ZDJ07}
\begin{barticle}[auto:STB|2014/02/12|14:17:21]
\bauthor{\bsnm{Zhu},~\bfnm{Y.}\binits{Y.}},
\bauthor{\bsnm{Dass},~\bfnm{S.~C.}\binits{S.~C.}} \AND
\bauthor{\bsnm{Jain},~\bfnm{A.~K.}\binits{A.~K.}}
(\byear{2007}).
\btitle{Statistical models for assessing the individuality of fingerprints}.
\bjournal{IEEE Transactions on Information Forensics and Security}
\bvolume{2}
\bpages{391--401}.
\end{barticle}
\bptok{imsref}%
\endbibitem

\end{thebibliography}
\end{document}